\documentclass[aps,twocolumn,twoside]{revtex4}
\usepackage{epsfig,amsmath,amssymb,dblfloatfix}
\usepackage{color}
\def\ket#1{|#1\rangle}

\newcommand{\ba}{\begin{eqnarray}}
\newcommand{\ea}{\end{eqnarray}}
\newcommand{\bmath}{\begin{mathletters}}
\newcommand{\emath}{\end{mathletters}}
\newcommand{\ban}{\begin{eqnarray*}}
\newcommand{\ean}{\end{eqnarray*}}

\newcommand{\bsub}{\begin{subequations}}
\newcommand{\esub}{\end{subequations}}

\begin{document}

%\title{Probing excited-state quantum phase transitions in cold trapped ion experiments}
\title{Probing excited-state quantum phase transitions with trapped cold ions}
\author{Marek Kucha{\v r}}\email{marek.kuchar@isibrno.cz} 
\affiliation{Czech Academy of Sciences, Institute of Scientific Instruments, Kr\'alovopolsk\'a 147, 61264 Brno, Czech Republic}
\affiliation{Charles University, Faculty of Mathematics and Physics, Institute of Theorertical Physics, Ke Karlovu 2, 12000, Prague, Czech Republic}

\author{Michal Macek}\email{michal.macek@isibrno.cz} 
\affiliation{Czech Academy of Sciences, Institute of Scientific Instruments, Kr\'alovopolsk\'a 147, 61264 Brno, Czech Republic}

\begin{abstract}
%We propose concrete protocols to realize and study quantum criticality due to excited-state quantum phase transitions (ESQPT) experimentally in presumably the simplest and most resilient system involving a single trapped ion oscillating in a radio-frequency Paul trap. The possibility of studying the ESQPT structure is based on the identification and detailed characterization of a specific class of excited states present between two ESQPT energies of the Extended Rabi Model (ERM) Hamiltonian. The framework proposed here allows for the methodical study of the critical scaling behaviors of witness observables for the various different criticalities of the ERM model. These can be examined by driving the system across the quantum criticalities by linearly changing the qubit-phonon coupling strength in time at different rates. A mapping of the theoretical control parameters of the ERM to the experimental parameters of a trapped ion setup is provided, and simulations are performed for values referencing existing state-of-the-art setups, addressing both unitary state evolutions and open system corrections.

We propose concrete protocols to realize quantum criticality due to excited-state quantum phase transitions (ESQPTs) experimentally in presumably the simplest and most resilient system involving a single trapped ion oscillating in a radio-frequency Paul trap. We identify a specific class of excited states of the Extended Rabi Model (ERM) Hamiltonian, which occur between two critical ESQPT energies of the model in its (anti)Jaynes-Cummings superradiant phase. Properties of these states motivate the definition of several ESQPT witness observables. We study their critical scaling behaviors as well as various distinct state evolutions by driving the system across the quantum criticalities by changing the qubit-phonon coupling strength linearly in time at different finite rates. A mapping of the theoretical control parameters of the ERM to the experimental parameters of a trapped ion setup is provided, and simulations are performed for values referencing existing state-of-the-art setups, addressing both unitary state evolutions as well as relevant open-system corrections.

\end{abstract}

%\pacs{} %21.60.Ev, 21.10.Re, 21.60.Fw, 24.60.Lz, 05.45.Mt}
% PACS, the Physics and Astronomy Classification Scheme.
%\keywords{Suggested keywords}
%Use showkeys class option if keyword display desired 
\date{\today}

\maketitle

\section{Introduction}\label{sec:Intro}
Phase transitions represent critical phenomena, in which equilibrium structure of the system undergoes abrupt (non-analytical) changes as a function of temperature or other control parameters and occur in a vast variety of dynamical systems across all scientific disciplines~\cite{SolePTbook}. Quantum phase transitions (QPTs) provide a specific example, in which the abrupt change occurs due to quantum fluctuations and affects the quantum ground state of the system at zero temperature~\cite{CarrBook,SachdevBook}, as noticed first by Hertz~\cite{Hertz76}. Numerous extensions of QPTs to non-equilibrium systems, including Dynamical Quantum Phase Transitions (DQPTs)~\cite{Heyl18}, or to criticalities involving the systems' excited spectrum, like the Excited-State Quantum Phase Transitions (ESQPTs)~\cite{Cejnar21,Monodromy,Cejnar08} have recently been proposed theoretically, while efforts for conceptual unification of the various approaches are under way~\cite{Corps25}. While DQPTs have already attracted considerable experimental interest, with the first experiments realizing the transverse-field Ising Hamiltonian in trapped ion chains~\cite{DQPTexpts1,DQPTexpts2}, the criticalities due to ESQPTs have so far been experimentally untouched. Perhaps closest to experimental realization so far is the proposal for detecting ESQPTs in spinor Bose-Einstein condensates~\cite{Feldmann21}. Profound changes induced by ESQPT quantum criticality, including advanced forms of quantum state manipulation and abrupt generation of various types of quantum entanglement can have implications for, and eventually practical use in emerging quantum technologies, especially metrology~\cite{Frerot18} and sensing~\cite{Beaulieu25,Gietka25}.

Individual trapped cold ions~\cite{Leibfried03} provide arguably the simplest and most controllable physical platform to realize quantum critical phenomena. Quite remarkably, as shown first by Hwang et al.~\cite{Hwang15}, quantum criticality can be realized by even a single quantum bit (qubit, realized by two internal states of the ion) interacting with a single continuous degree of freedom---in this case the ion's one-dimensional motional mode (phonon). Signatures of the quantum criticality are in this case realized not in the infinite-size thermodynamic limit, but in the limit of infinite ratio between the qubit and the phonon excitation energies~\cite{Hwang15}. Interaction of the qubit with the phonon is described by the Rabi (R), or in simpler cases by the Jaynes-Cummings (JC), or anti-Jaynes-Cummings (aJC) Hamiltonians.

Experimental realization of the Rabi interaction with sufficient 'strength' (sufficiently strong coupling between the qubit and the phonon degrees of freedom) to address the criticalities is however not straightforward. In typical cold-ion setups using radio-frequency Paul traps combined with several cooling and driving laser fields, the high-intensity laser fields lead to breaking of the Lamb-Dicke approximation and introduce higher-order qubit-phonon interactions beyond the Rabi interaction. A practical protocol enabling to generate a `synthetic' Rabi Hamiltonian in regimes of strong interaction (ultra/deep strong interaction, USC/DSC), however without addressing the possible quantum criticalities, was identified recently by Pedernales et al. in~\cite{Pedernales16}. The protocol uses a combination of driving lasers suitably detuned from the so-called red and blue motional sidebands of the ion. This protocol was later explicitly used for the experimental demonstration of the QPT in the ground state of the Rabi Hamiltonian using an $^{171}$Yb$^+$ ion in a Paul trap~\cite{Duan21} driven using relatively slow, and linear-in-time protocols of changing the qubit-phonon interaction parameters.
%~\footnote{The authors used a common, although somewhat misleading terminology for the `quench'...}.      

In this paper, we propose experimentally feasible driving protocols to realize the ESQPT quantum criticalities in the spectrum of a single cold trapped ion and identify classes of criticality-related excited states with possible applications in quantum sensing~\cite{FilipManyFock}. We work out in detail the expected behaviors of several ``ESQPT-witnesses'' observables under these driving protocols. Attention is paid to concrete technical parameters of state-of-the-art experimental setups, in particular the $^{40}$Ca$^+$ ion trap operated at ISI Brno.

The structure of the paper is as follows: In Sec.~\ref{sec:2}, we map the control parameters of the extended Rabi Hamiltonian of Ref.~\cite{Stra21} to experimental parameters of a typical radio-frequency Paul trap and the driving laser frequencies; in Sec.~\ref{sec:3}, we review the QPT and ESQPT structure of the extended Rabi model Hamiltonian; in Sec.~\ref{sec:4} we highlight a specific class of excited states of the time-independent Hamiltonian based on which we define ESQPT-witness observables; in Sec.~\ref{sec:5} we overview different time-dependent driving protocols and propose observables suitable as experimental witnesses of the QPTs and ESQPTs in the driven system under unitary evolution; in Sec.~\ref{sec:ExptParamMaster}, we consider the non-unitary (open-system) evolution with experimentally relevant sources of dissipation within an appropriate master equation framework, and compare the Rabi flops expected to be observed in states, which experienced different QPT/ESQPT phases; finally in Sec.~\ref{sec:Concl}, we bring the conclusions. 

%{\color{blue} Marek, please use blue for your inputs...okay, i will}

\section{Extended Rabi Hamiltonian in Trapped Cold Ion Experiments}\label{sec:2}
% \begin{itemize}
%     \item{Trap and its parameters - rf. freq, lasers, ...}
%     \item{Rabi and JC/aJC interaction from the general El-Mag Hamiltonian - Lamb-Dicke parameter, RWAs,..(a'la Pedernales - why is it necessary?)}
%     \item{Extended Rabi H of Stransky-Cejnar-Filip and its mapping to the experimental system (application of Pedernales approach)}
% \end{itemize}
In order to quantize the motion of a charged ion in the radio-frequency Paul trap, we adapt the reference oscillator representation \cite{Leibfried03}. The Heisenberg-picture equation of motion admits a Floquet-type solution with components oscillating at the secular frequency $\nu$, superimposed with multiples of the trap driving radio frequency $\omega_{\rm rf}\gg\nu$, representing the micromotion. Micromotion is usually neglected entirely, but it may be included at lowest order as an effective weakening of the coupling.

The ion's two-level structure with energy separation $\hbar\omega_0$, valid under near-resonant driving, can be coupled to the motional degree of freedom by applying external laser field \cite{james, GardinerZollerBooks}. The qubit-motion interaction can be expanded into individual $\omega_j$ components in terms of Rabi frequencies $\Omega_j$, which represent the magnitudes of the qubit-state transition matrix elements. The interaction Hamiltonian becomes:
\begin{equation}\label{1}
    \hat{H}^{(i)}(t)=\hbar\hat{\sigma}_x\sum_j\Omega_je^{i\phi_j}e^{i(\mathbf{k}_j\cdot\hat{\mathbf{r}}-\omega_j t)}\; +\rm{H.\,c.\:,}
\end{equation}
\noindent with $\hat{\bm{\sigma}}_i$ denoting the Pauli matrices, $\mathbf{k}_j$ the wave vector of the $j$-th coupling field component, and the sum running over all frequencies present, which must satisfy \mbox{$|\omega_0-\omega_j|\ll\omega_0$}.

Rotating to the interaction picture with respect to the sum of the two-level and motional non-interacting Hamiltonians results in two pairs of terms. Rapidly oscillating terms $\propto e^{\pm i(\omega_j+\omega_0)}$ can be neglected via the first rotating-wave approximation (RWA), and we are left with
\begin{equation}\label{2}
\begin{aligned}
\hat{H}_{\rm int}^{(1)}(t)=\frac{\hbar}{2}\hat{\sigma}_+\sum_j \Omega_j&e^{i\left[\phi_j-(\omega_j-\omega_0)t+\eta_j\left(\hat{a}(t)+\hat{a}^\dagger(t)\right)\right]}\\
&\quad\quad\quad\quad\quad\;+\rm{H.\,c.}\;,
\end{aligned}
\end{equation}
\noindent where $\hat{a}(t)=\hat{a}e^{-i\nu t}$ is the bosonic annihilation operator and $\eta_j=k_j\sqrt{\hbar/(2m\nu)}$ is the Lamb-Dicke parameter. It represents the maximal shift of the coupling-field phase over the extent of the ion's trajectory and is usually considered small, allowing for an expansion of the exponential. Additionally, the restriction placed on the coupling-field frequencies, which stems from the first RWA, allows us to approximate $\eta_j\approx\eta\equiv\eta(\omega_0)$.

In the Lamb-Dicke regime, where we neglect $O(\eta^2)$ terms in \eqref{2}, one can reduce the interaction to various known models through an optimal choice of coupling frequencies $\omega_j$ and subsequent omission of the remaining oscillating terms via a second round of RWAs. Examples include the carrier resonance \mbox{($\omega_1=\omega_0$),} the Jaynes-Cummings and anti-Jaynes-Cummings models \mbox{($\omega_1=\omega_0\mp\nu$}, corresponding to the first red and blue sideband resonances, respectively), and the Rabi model \mbox{($\omega_1=\omega_0-\nu$, $\omega_2=\omega_0+\nu$)}, which requires bichromatic coupling. Outside the Lamb-Dicke regime, one can even realize effective motional squeezing \cite{Huerta} by coupling to both second sidebands \mbox{($\omega_1=\omega_0-2\nu$, $\omega_2=\omega_0+2\nu$)}, provided we initially prepare the qubit in an equally weighted superposition, e.g. in  an eigenstate of the $\sigma_x$ operator.

However, in order to reach the strong-coupling regimes necessary to study the QPT and ESQPT criticalities, we need to adapt the detuned-coupling approach proposed in \cite{Pedernales16}, which has been experimentally realized to examine the ground-state QPT in the standard Rabi model \cite{Duan21}. In the case of the Rabi model, this approach involves detuned bichromatic coupling to the first sidebands via frequencies $\omega_1=\omega_0-\nu+\delta_r$ and $\omega_2=\omega_0+\nu+\delta_b$, where $\delta_{r,b}\ll\nu$ represent detunings from resonance. The second RWA reduction, assuming constant phase matching, yields
\begin{equation}\label{3}
    \hat{H}_{\rm int}^{(2)}(t)=\frac{\hbar\eta}{2}\hat{\sigma}_+\left(\Omega_1\hat{a}e^{-i\delta_r t}+\Omega_2\hat{a}^\dagger e^{-i\delta_b t}\right)+\rm{H. c.}
\end{equation}
\noindent We can rotate this Hamiltonian into a time-independent one if we define the artificial non-interacting Hamiltonian
\begin{equation}\label{4}
    \hat{H}_0=-\frac{\hbar(\delta_r+\delta_b)}{4}\hat{\sigma}_z+\frac{\hbar(\delta_r-\delta_b)}{2}\hat{a}^\dagger\hat{a}\,.
\end{equation}

\noindent By performing this incomplete rotation back to the Schr\"odinger picture and not imposing the condition of equal Rabi frequencies for the individual sideband drivings, we can map the trapped-ion system onto the extended Rabi model (ERM)~\cite{Stra21}
\begin{equation}\label{5}
    \hat{H}=\hat{H}_0+\hbar\Lambda\left[\frac{1+\delta}{2}\left(\hat{\sigma}_+\hat{a}+\hat{\sigma}_-\hat{a}^\dagger\right)+\frac{1-\delta}{2}\left(\hat{\sigma}_+\hat{a}^\dagger+\hat{\sigma}_-\hat{a}\right)\right],
\end{equation}
\noindent where 
\begin{equation}\label{eq:Lambdadelta}
    \Lambda=\eta(\Omega_1+\Omega_2)/2\;,\quad  \delta=(\Omega_1-\Omega_2)/(\Omega_1+\Omega_2)\,.
\end{equation}

We further adjust the model parameters in \eqref{5} in order to provide a clearer interpretation when analyzing the criticalities present. By demanding $(\delta_r+\delta_b)<0$ and $\delta_r>\delta_b$, we obtain the naturally expected behavior in $\hat{H}_0$ and can define a characteristic energy scale 
\begin{equation}\label{eq:eps}
\varepsilon=\frac{\hbar}{2}\sqrt{\delta_b^2-\delta_r^2}\,,
\end{equation}
the effective system size \cite{Hwang15}\cite{Stra21} 
\begin{equation}\label{eq:Del}
\Delta=(\delta_b+\delta_r)/(\delta_b-\delta_r)\,,
\end{equation}
and the rescaled coupling strength 
\begin{equation}\label{eq:lam}
    \lambda=2\Lambda/\sqrt{\delta_b^2-\delta_r^2}\,.
\end{equation}
In terms of the spin-$1/2$ operators $\hat{J}_i=\hat{\sigma}_i/2$, we arrive at the final version of the ERM Hamiltonian 
\begin{equation}\label{6}
\begin{aligned}
    &\quad\quad\quad\hat{h}=\frac{1}{\varepsilon\sqrt{\Delta}}\hat{H}=\hat{J}_z+\frac{1}{\Delta}\hat{a}^\dagger\hat{a}\,+\\
    &\frac{\lambda}{\sqrt{\Delta}}\left[\frac{1+\delta}{2}\left(\hat{J}_+\hat{a}+\hat{J}_-\hat{a}^\dagger\right)+\frac{1-\delta}{2}\left(\hat{J}_+\hat{a}^\dagger+\hat{J}_-\hat{a}\right)\right]\,.
    \end{aligned}
\end{equation}
\noindent Here $\delta\in(-1,\,1)$ interpolates from the aJC regime ($\delta=-1$), through the standard Rabi regime ($\delta=0$) to the JC regime ($\delta=1$), and corresponds precisely to the Hamiltonian of Eq.~(5) in~\cite{Stra21} with $\mu=0$ (the parity-conserving case). Required values of the parameters of Eqs.~(\ref{eq:Lambdadelta})-(\ref{eq:lam}) and the conditions for their realization, demonstrating in particular when the ESQPT criticalities are experimentally feasible, are discussed in Appendix~\ref{App:C}.

\section{QPT and ESQPT in systems involving a single trapped ion }\label{sec:3}
% Time independent systems with fixed $\lambda, \delta$.
% \begin{itemize}
%     \item{Semiclassical underpinning ot the QPT/ESQPT analysis - classical Hamiltonian}
%     \item{Phase diagram}
%     \item{Classical Hamiltonians - Figure - JC/aJC-Mex.Hat, Rabi, $\delta=0$}
%     \item{Level dynamics}
%     \item{Wigner functions}
%     \item{Entanglement - von Neumann entropy}
%    % \item{Finite size scaling of the "strength function" in the ESQPT-phase (between $E_{sad}$ and $E_{loc.max.}$); not just "finite-size" but "energy scaling" with respect to ($E - E_{sad}$)}

%     \item{Observables - $n$, $\vec{J}$}, ...   
% \end{itemize}

Key signature of ESQPT criticalities in systems with $f=1$ degree of freedom---such as described by the Hamiltonian~\eqref{6}---are divergences (or discontinuities/jumps) in the semiclassical energy level density $\rho(E,\lambda)$ at specific excitation energy $E_\mathrm{crit}$ and/or at critical values of the control parameters $\lambda_\mathrm{crit}$. For a comprehensive review and classification of ESQPTs, see~\cite{Cejnar21}. 
%Types of ESQPTs are classified based on the (i) number of degrees of freedom $f$ and (ii) the types of different unstable equilibria (local minima, local maxima, saddle points, etc...) of the semiclassical Hamiltonian $h$ of the model, leading to specific non-analyticity of $\rho(E,\lambda)$, most pronounced in low-$f$ systems. 
The model~\eqref{6} allows to control and study the finite size scaling and the type of criticality~\cite{CarrBook,SachdevBook}, which can be adjusted in state-of-the-art experimental trapped ion setups via Eq.~\eqref{eq:Del}. 
Before analyzing concrete ESQPT-witness observables, dependent on $\Delta$ of Eq.~\eqref{eq:Del}, let us for convenience recapitulate the phase structure of the model derived in~\cite{Stra21}: 

% (the following section formated in italics should be reformulated in a complete but as simple as possible form, as it contains information derived already by~\cite{Stra21} - MM will do it...

The semiclassical Hamiltonian $h(\lambda,\delta)$ corresponding to the extended Rabi model (\ref{6}) is obtained if 
%{\color{blue} The analysis of (ES)QPTs relies on a known semiclassical approach \cite{Stra21}.
we expand the bosonic operators as $\hat{a}=(m\nu \hat{x}+i\hat{p})/\sqrt{2m\hbar\nu}$ and rescale $\hat{x}'=\hat{x}\sqrt{m\nu/(\Delta\hbar)}$ and \mbox{$\hat{p}'=\hat{p}/\sqrt{\Delta m\hbar\nu}$.} This leads to the canonical commutation relation \mbox{$[\hat{x}',\hat{p}']=i/\Delta$.} Therefore, the motional classical limit can be realized in the $\Delta\to\infty$ limit, where the coordinate and momentum operators may be treated as classical phase-space coordinates. A semiclassical counterpart of \eqref{6} is then given by
\begin{equation}\label{7}
 h(x',p',m')=\frac{x'^2+p'^2}{2}+m'\sqrt{2\lambda^2(x'^2+\delta^2p'^2)+1}\,, 
\end{equation}
\noindent where $m'=\pm1/2$ are the eigenvalues of the projected quasispin operator $\hat{J}_z'=\mathbf{n}\cdot\hat{\mathbf{J}}\,$, which acts as an approximate integral of motion at low excitation energies \cite{Armando}. The parameter $m'$ therefore distinguishes two classically non-interacting subsystems. When searching for stationary points, one quickly concludes that the $m'=+1/2$ subsystem has only a global minimum at $(x',p')=(0,0)$ for arbitrary $\lambda\,,\delta$ and therefore does not exhibit any criticality. However, energy levels corresponding to this subsystem will still appear in any finite-size spectrum at $e>1/2$ and resemble a deformed harmonic oscillator. The structure and character of the stationary points present in the \mbox{$m'=-1/2$} subsystem dictate the phase structure of the system.
 Thus, from the perspective of the ESQPT analysis, the ERM restricted to $m'=-1/2$ is a one-dimensional ($f=1$) system. 

\begin{figure}[b!]
\begin{center}
    \includegraphics[width=\linewidth]{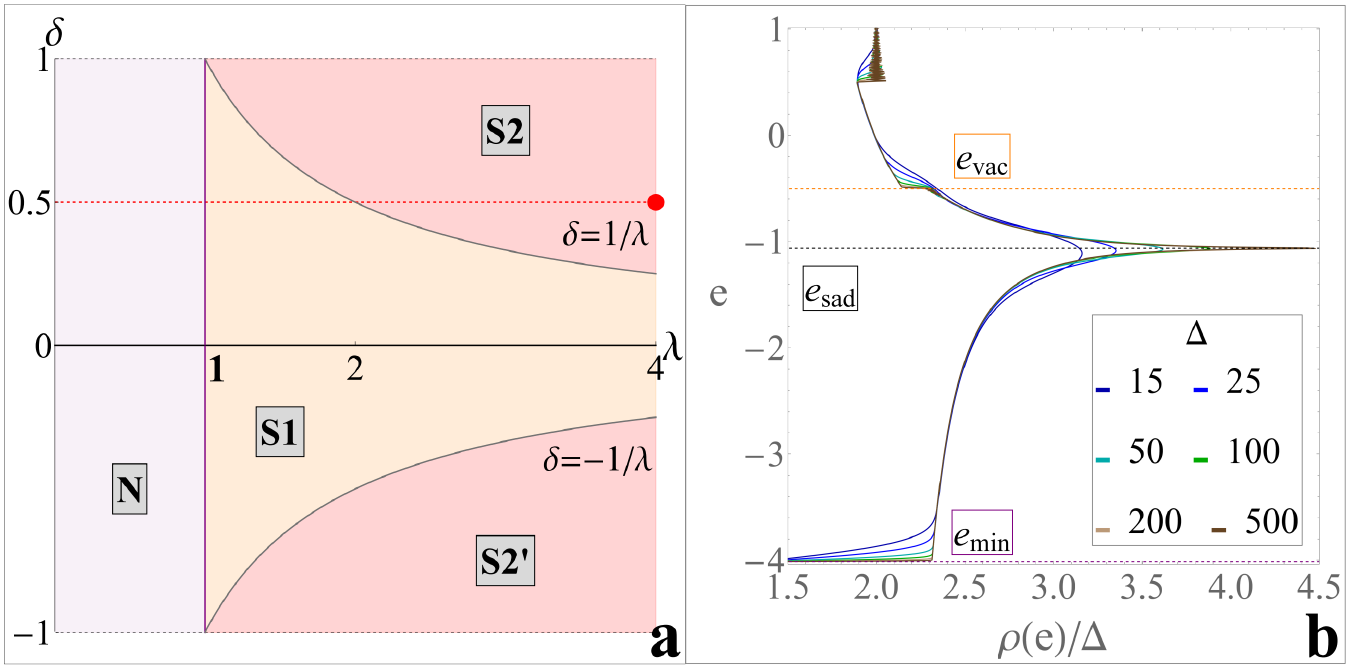}
\end{center}
\caption{(a) Different qubit-phonon interaction phases of the semiclassical ERM Hamiltonian~\eqref{7} in the $(\lambda, \delta)$ plane. The individual colored regions represent areas with different phonon-vacuum properties. The red horizontal line at $\delta=0.5$ represents our choice of typical parameter-space path. Red point at $(\lambda,\,\delta)=(4,\,0.5)$ corresponds to parameters, for which we plot the Gaussian-smoothed densities of states (b) for different system sizes $\Delta$. Horizontal dashed lines in (b) indicate local maximum (orange), saddle point (black) and global minimum (purple) of the classical  Hamiltonian.} 
\label{fig:1}
\end{figure}

While the ground state QPTs in~\eqref{6} and \eqref{7} involve transitions between the normal (N) and superradiant (S) phases, occurring at $\lambda_c \equiv 1$ (independent of $\delta)$, the ESQPTs provide a richer set of options, with possible coexistence of two types of superradiant phases in the excitation spectrum: the Rabi (S1) and Jaynes-Cummings (S2) superradiant phases. As done first in~\cite{Stra21}, these are best characterized by stability of the phonon vacuum state, which changes along two additional critical lines $ \lambda_0 \equiv 1/|\delta|$, for positive/negative $\delta$, respectively. We replot `the ESQPT phase diagram' for convenience in Fig.~\ref{fig:1} (panel a): In a fixed interaction-type regime ($\delta=\mathrm{const.}$), we can classify the motion-qubit coupling strength phases as follows:
\begin{itemize}
    \item $\lambda\in[0,1)$ - \textit{Normal (weakly interacting) phase N}: The only stationary point is the global minimum at the origin $(x',p')=(0,0)$ corresponding to the bosonic vacuum at energy $e_\mathrm{vac} \equiv -1/2$.
    \item $\lambda\in(1,1/|\delta|)$ - \textit{First (Rabi) superradiant phase S1}: Two stationary points located at \mbox{$(\pm x_c',p')\equiv\left(\frac{\pm\lambda}{\sqrt{2}}\left(1-\lambda^{-4}\right)^{1/2},0\right)$} represent degenerate global minima at energy \mbox{$e_{\rm min}=-\frac{1}{4}(\lambda^2+\lambda^{-2})$}. The origin $(x',p')=(0,0)$ becomes a saddle point accompanied by a logarithmic divergence of the semiclassical density of states~\cite{Cejnar21} at the energy $e_\mathrm{vac}$, which represents the (only) ESQPT criticality in the S1 phase. 
    \item $\lambda\in(1/|\delta|,\infty)$ - \textit{Second (Jaynes-Cummings) superradiant phase S2}: Two degenerate global minima are given by the same expressions as in the S1 phase, while two energetically degenerate saddle points arise at \mbox{$(x',\pm p_c')\equiv\left(0,\frac{\pm\lambda\delta}{\sqrt{2}}\left[1-(\lambda\delta)^{-4}\right]^{1/2}\right)$} and lead to a logarithmic divergence of the semiclassical density of states at energy \mbox{$e_{\rm sad}(\lambda, \delta)\equiv-\frac{1}{4}\left[(\lambda\delta)^2+(\lambda\delta)^{-2}\right]$;} For $|\delta| < 1$, this energy  represents the lower-lying ESQPT criticality in the S2 phase. For $|\delta| = 1$ (pure JC or aJC Hamiltonian), the energies $e_{\rm sad}$ and $e_{\rm min}$ coalesce [mexican-hat form of~\eqref{7} related to U(1) symmetry of~\eqref{6}], and this ESQPT vanishes. The origin turns into a local maximum, leading to a non-analytic drop in the density of states~\cite{Cejnar21} at $e_\mathrm{vac}$, which represents the higher-lying ESQPT criticality in the S2 phase. An example is given in Fig.~\ref{fig:1} (b) for $\delta = 0.5$ and $\lambda = 4$, where the development of the non-analytic behaviors at the two ESQPT energies, $e_\mathrm{vac}$ (jump) and $e_\mathrm{sad}$ (log-divergence) with increasing $\Delta$ is seen. 
\end{itemize}
 
 As the semiclassical Hamiltonian~\eqref{7} is $\delta \leftrightarrow -\delta$ symmetric, so is the ESQPT phase structure. To eventually account for finite-size differences in the respective quantum systems~\eqref{6}, we denote the anti-Jaynes-Cummings superradiant phase by S2', in line with Ref.~\cite{Stra21}. Detailed properties of the system in the S2 and S2' phases are further discussed in the Appendix A.
 
\begin{figure}[t!]
\begin{center}
    \includegraphics[width=\linewidth]{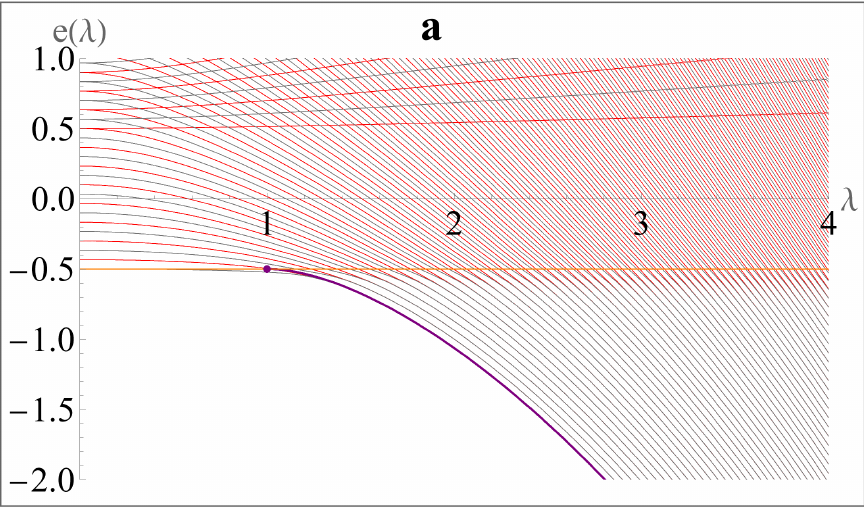}
    \includegraphics[width=\linewidth]{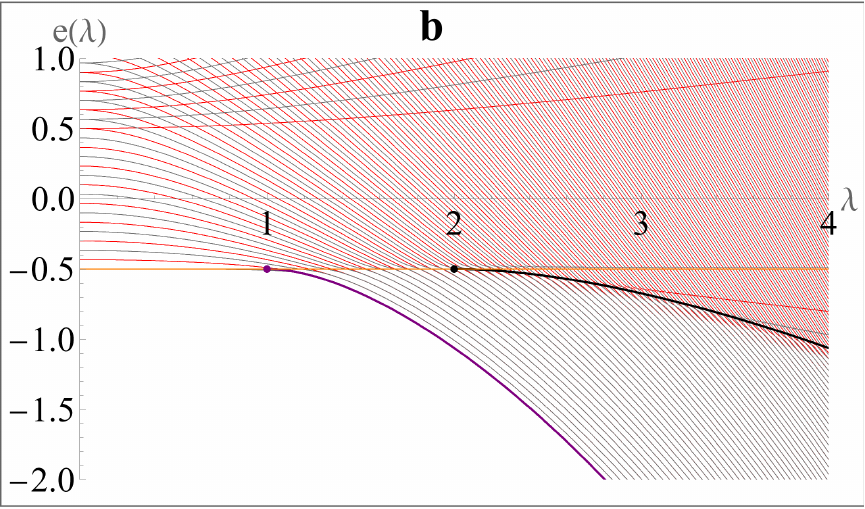}
    \includegraphics[width=\linewidth]{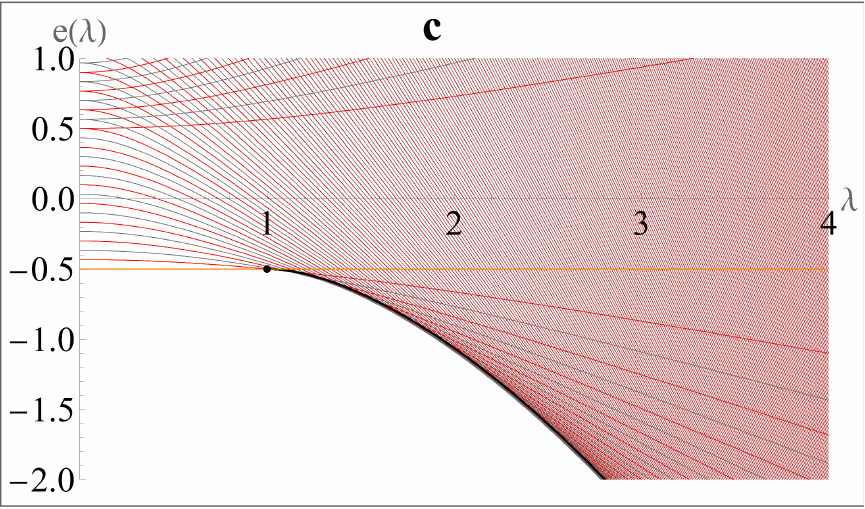}
\end{center}
\caption{Level dynamics of the ERM Hamiltonian~\eqref{6}, with system size $\Delta=15$ as a function of interaction strength $\lambda$ for (a) $\delta=0$, (b) $\delta=0.5$ and (c) $\delta=1$. Red levels indicate eigenstates with negative parity and gray levels indicate eigenstates with positive parity with respect to the total number of excitations. The orange line represents the energy of the non-interacting ground state. The purple dot at $\lambda_c$ indicates the ground-state QPT, and the subsequent purple line $e_{\rm min}$ represents the ground-state energy. In (b), the black dot at $\lambda_0$ and the subsequent black line $e_{\rm sad}$ indicate the second ESQPT energy, resulting from the transition to the second superradiant phase S2. In panel (c) $\lambda_0 = \lambda_c$ and $e_{\rm sad} = e_\mathrm{min}$.
} 
\label{fig:2}
\end{figure} 

%On the basis of these results, we acquire complete information concerning the (ES)QPT structure. At $\lambda_c=1$, the system undergoes a ground-state QPT, as a result of which the global minimum at energy \mbox{$e_{\rm min}=-\frac{1}{4}(\lambda^2+\lambda^{-2})$} no longer coincides with the bosonic vacuum at energy $e_\mathrm{vac} = -1/2$. In the generic cases $\delta\notin\{0,\pm1\}$, the non-interacting vacuum energy $e_{\rm vac}=-1/2$ corresponds to an ESQPT exhibiting level bunching up to $\lambda_0=1/|\delta|$, where the density of states suddenly drops. Beyond $\lambda_0$, an additional ESQPT energy forms at $e_{\rm sad}(\lambda, \delta)=-\frac{1}{4}\left[(\lambda\delta)^2+(\lambda\delta)^{-2}\right]$}.

\section{Witnesses of ESQPTs: Time-independent Hamiltonian }\label{sec:4}
There are ongoing efforts to observe precursors of non-analytic behaviors of $\rho(e,\lambda)$ at ESQPT critical values experimentally \cite{Chavez-Kerr,Champion}, however, we focus here on identifying concrete ESQPT witness observables for experiments with trapped ion setups. Our proposed system is characterized by the critical values of $e_\mathrm{crit}\in\{e_\mathrm{sad},\,e_\mathrm{vac}\}$ and \mbox{$\lambda_\mathrm{crit}\in\{\lambda_c,\lambda_0\}$}. Thus, we aim to utilize the Hamiltonian of Eq.~\eqref{6} with finite effective system size $\Delta < \infty$, Eq.~\eqref{eq:Del}, generalizing the protocols used in the first experimental demonstration of the ground-state QPT in the pure Rabi model in~\cite{Duan21}. Their work showed that a trapped ion setup is able to withstand environmental effects sufficiently to observe QPT precursors in the natural observables $\hat{n}$, $\hat{J}_z$ and we extend these principles to the ESQPTs.

In Fig.~\ref{fig:2}, we show all possible scenarios for development of QPT and ESQPT criticalities of Eq.~\eqref{6}. The Figure displays $\lambda$-dependent `level dynamics' of the pure Rabi model with $\delta = 0$ (panel a), the ERM with $\delta=0.5$ (b), and the pure Jaynes-Cummings model (c) along the same range of $\lambda\in[0,4]$ and with a matching (moderate) system size $\Delta=15$. Spectra in (a) and (b) enter the first (Rabi) superradiant phase S1 at a finite-size corrected $\tilde{\lambda}_c\gtrsim\lambda_c$, as a result of which opposite-parity mergers occur approximately at the ESQPT energy $e_\mathrm{vac}$, accompanied by a rapid drop of the ground-state energy with increasing $\lambda$ due to the QPT. The JC spectrum (c) lacks the S1 phase and enters at $\tilde{\lambda}_c\gtrsim\lambda_c$ the S2 superradiant phase, instead. The ERM spectrum (b) displays both the S1 and S2 superradiant phases: the second ESQPT is formed for $\tilde{\lambda}_0\gtrsim\lambda_0$ at energy $e_{\rm sad}(\lambda, \delta)$. The corresponding spectrum contains only S1-type excitations for $e < e_{\rm sad}(\lambda, \delta)$ and S2-type of excitations for $e > e_{\rm sad}(\lambda, \delta)$.
%To emphasize the rich parametric structure of interaction in the ERM compared to the standard Rabi model, we show the regions of different field-vacuum character in the $(\lambda,\delta)$-plane diagram in Fig.~1. The diagram is divided into three regions corresponding to the phase classification described previously. While the $\delta\in\{0,\pm1\}$ regimes span only two distinct phases, any intermediate value can reach an arbitrary region of the diagram via tuning of the coupling strength.}
%We show that the spectral interval bounded by the two stationary points in the excited spectrum becomes crucial for examining the dynamical signatures of the ESQPTs.  % Statement of this kind could be used elsewhere (e.g. in Conclusions); here it somewhat disrupts the flow of the text.
Note that the Hamiltonian \eqref{6} with general $\delta$ possesses a discrete $\mathbb{Z}_2$ symmetry with respect to the parity operator \mbox{$\hat{P}=(-1)^{\hat{n}+\hat{J}_z+1/2}$.} Additionally, in the $\delta=\pm1$ cases (panel c), the symmetry group extends to a continuous $U(1)$, allowing for a simple analytic solution and a large number of direct crossings of the same parity in the coupling-dependent level dynamics $e_n(\lambda)$~\cite{Braak11,dingshun}. In contrast, the level crossings of the same parity shown in Fig.~\ref{fig:2} (a) and (b) are all avoided. The only direct crossings allowed by symmetry are those of opposite parity states. % and they occur at exceptional points in the spectrum \cite{Chen21}. % MM note: This statement is not necessary; moreover 'exceptional point' is a technical term denoting specific non-Hermitian degeneracies, which we do not discuss here. 

\begin{figure}[b!]
  \centering
  \includegraphics[width=\linewidth]{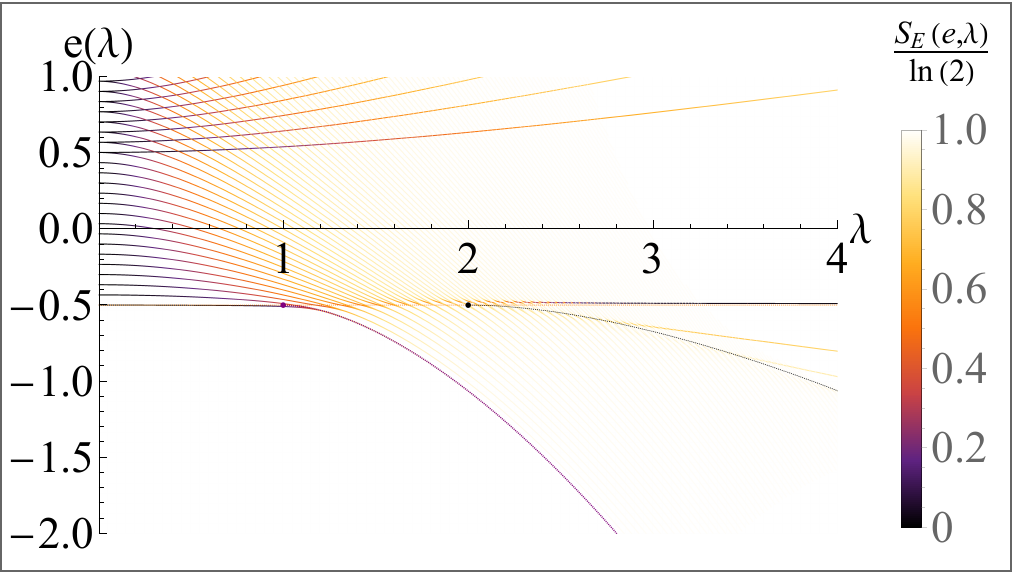}
  \caption{Level dynamics as in Fig. 2 (b), but color-coded by the qubit-motion entanglement entropy $S_E$~\cite{GardinerZollerBooks}. $S_E$ is normalized by its maximal value of $\ln(2)$.}
  \label{fig:entr}
\end{figure}

Key difference between the ESQPT scenarios in panels (a) and (b) of Fig.~\ref{fig:2}, is the appearance of a specific class of eigenstates, denoted hereafter as {\it S2 emergent states}, for $\lambda \gtrsim \tilde{\lambda}_0$ in the S2 phase between the two ESQPT energies $e \in [e_\mathrm{sad}(\lambda, \delta), e_\mathrm{vac}]$: The S2 emergent states correspond to the subset of energy levels with relatively low negative $\tfrac{\partial e}{\partial \lambda}$ (for both parities), which undergo sharply avoided level-crossings with the remaining subset of `background levels' with relatively much steeper negative $\tfrac{\partial e}{\partial \lambda}$.  The class of S2 emergent states involves in particular the slightly modified phonon vacuum ($\tfrac{\partial e}{\partial \lambda} \approx 0$ for $\lambda \gtrsim \tilde{\lambda}_0$), stabilization of which in the S2 phase was observed in~\cite{Stra21}.
Let us note in passing that for $\delta < 0$ an analogous class of {\it S2' emergent states} exists between $e \in [e_\mathrm{sad}(\lambda, \delta), e_\mathrm{vac}]$ for $\lambda \gtrsim \tilde{\lambda}_0$, however, it does not contain the stabilized vacuum (cf. Appendix~\ref{App:A}). The coexistence of the S2 emergent states and the remaining `background states' is actually most clearly visible in the $\lambda \gtrsim \tilde{\lambda}_c$ spectrum of the pure JC model (panel c): Here however the lower energy bound for their occurrence does not constitute an ESQPT, as it coincides with the ground state energy itself. Furthermore, we note that the S2 emergent states also constitute the only negative energy eigenstates without (nearly) maximal qubit-motion entanglement in the $\lambda\geq\tilde{\lambda}_0$ and $e \in [e_\mathrm{sad}(\lambda, \delta), e_\mathrm{vac}]$ spectral region. We show this in Fig. \ref{fig:entr}, which displays the 'spectral dynamics' from Fig. 2 (b) color-coded by the bipartite entanglement entropy~\cite{GardinerZollerBooks}. 

In both Figs.~\ref{fig:2},$\,$\ref{fig:entr}, we observe states at $e>1/2$, which relatively slowly gain energy with increasing $\lambda$ and undergo the same avoided crossings dictated by parity. However, these eigenlevels should not be mistaken for the S2 emergent states, since we see that their existence is not limited by a specific $(\lambda,\delta)$ region. In fact, these states can be attributed to the $m'=+1/2$ subsystem of \eqref{7}, which will be evident from the analysis below.

 \begin{figure}[t!]
\begin{center}
    \includegraphics[width=1.008\linewidth]{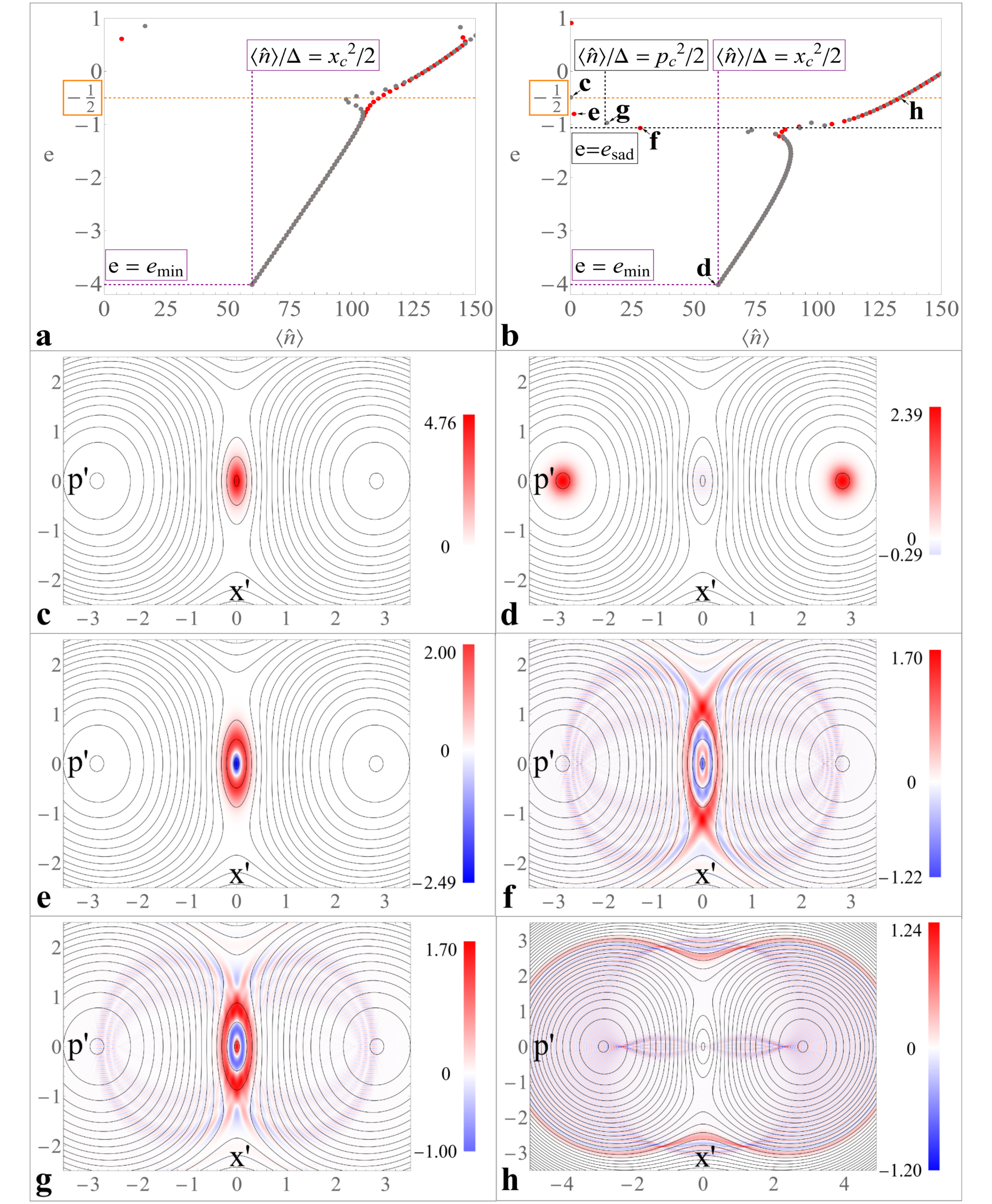}
\end{center}
\caption{Expectation values of the phonon-number operator $\langle\hat{n}\rangle$ in eigenstates of the Hamiltonian \eqref{6} with $\Delta=15$, $\lambda=4$ and $\delta=0$ (panel a)  and $\delta=0.5$ (b) with six selected states' Wigner functions in panels (c) - (h). The vertical axes in (a) and (b) display the eigenvalues, and the horizontal axes show the mean number of phonons in the corresponding parity-distinguished eigenstate (gray for positive parity, red for negative parity). The orange dashed horizontal line marks the non-interacting ground-state energy [semiclassical saddle point in (a) and local maximum in (b)], the purple horizontal line marks the ground-state energy (semiclassical minimum) and the purple vertical line marks the corresponding phonon number. The black horizontal line in (b) marks the second ESQPT energy (semiclassical saddle point) and its vertical counterpart likewise displays the corresponding phonon number. States shown in (c) - (h) are indicated in (b), and their Wigner functions are displayed along with contours of the classical Hamiltonian $h(x',p',-1/2)$ \eqref{7}.} 
\label{fig:5}
\end{figure} 
%{\bf Toward `intuitive interpretation of S2 emergent states: show and briefly discuss here the von Neumann entanglement entropy~\cite{GardinerZollerBooks} - emergent states weakly entangle motional and qubit DoFs...Continue with Peres and Wigner...}
 
Criticalities of the system translate to interesting behavior of the expectation values of specifically chosen observables, displayed in the so-called Peres lattices~\cite{Peres}. Natural observables of choice correspond to operators invariant with respect to the evolution-picture rotations performed in the ERM mapping procedure, namely $\hat{n}$, $\hat{J}_z$ or functions thereof. Furthermore, since the (ES)QPTs have been found as criticalities of a classical phase-space Hamiltonian, we also discuss selected states' Wigner functions, defined as
\begin{equation}
    W_\rho(x',p')=\frac{\Delta}{\pi}\int_\mathbb{R}\langle x'-y'|\hat{\rho}_m|x'+y'\rangle e^{2i\Delta y'p'}dy'\,,
\end{equation}
\noindent where $\hat{\rho}_m=\text{Tr}_q|\psi\rangle\langle\psi|$ is the reduced density operator of the motional degree of freedom.%Since the phonon-basis decomposition of states will later be important when modeling post-driving state diagnostics, 

In Fig.~\ref{fig:5}, panels (a) and (b), we show in particular the Peres lattice of the boson number operator $\hat{n}$ for $\delta = 0$ (panel a) and $\delta = 0.5$ (b). The set of points $\{(\langle \hat{n}\rangle_i, e_i)\}_{i=0}^N$, where $i$ runs over the system's eigenstates $|\psi_i\rangle$, is in both cases plotted for $\lambda = 4$, i.e. in the S1 for (a) and S2 for (b), and for a system size $\Delta = 15$.
Examples of significant Wigner functions of the ERM eigenstates, for the same parameters as in the Peres lattice (panel b), are shown in panels (c-h), color-coded and shown on top of the contour-lines of the corresponding classical Hamiltonian~\eqref{7}. 
%Both systems are already beyond $\tilde{\lambda}_c$, and therefore their ground-state energy $e_{\rm min}$ lies below the non-interacting ground-state $e_{\rm vac}=-1/2$. T
The Wigner function for the ground-state (d) shows that the state is equally split into the double-well minima determined by the classical Hamiltonian, and from the Peres lattice we see that $\langle\hat{n}\rangle_{\rm gs}/\Delta=x_c^2/2$ is satisfied. 
Since the $\delta=0.5$, $\lambda = 4$ system is already deep within S2, $e_{\rm vac}$ corresponds to the local maximum of~\eqref{7} and $e_\mathrm{sad}$ is a saddle point of~\eqref{7}. We find specific Peres lattice points in (b) in the $(e_{\rm sad}, e_{\rm vac})$ interval with significantly lower values of $\langle \hat{n}\rangle$ than their neighbors in the spectrum; an approximate bound for which can be determined as $\langle\hat{n}\rangle\leq\langle\hat{n}\rangle_{\rm sad}=\Delta p_c^2/2$ (shown in b, explained in Appendix A). These correspond precisely to the S2 emergent states bound by the semiclassical maximum~\footnote{The S2 and S2' excited phases for $|\delta| \neq 1$ in fact display remnants of U(1) symmetry of the (a)JC Hamiltonians, endowed with the unusual kinetic term $H \propto -p^2$ locally around $x=p=0$ related to the (locally) Mexican-hat-form of the semiclassical (a)JC Hamiltonians of Eq.~\eqref{7}. A more rigorous interpretation possibly in terms of quasi- and partial dynamical symmetries is an interesting open question~\cite{MacLev14}.}. Their Wigner functions are shown in panels (c), (e) and (g), in order of decreasing energy starting from the stabilized vacuum. They approximately follow the contours of the classical Hamiltonian around the $x=p=0$ local maximum, which are elongated in the $p$-direction. The Wigner functions of the S2 emergent states are thus squeezed in the direction of the $x$-axis. Note that in (a), the $e_{\rm vac}$ ESQPT is a saddle point and therefore shows a behavior similar to $e_{\rm sad}$ in (b), however the absence of the second ESQPT point forbids any additional structure to form in (a) and we show it here for brief comparison. Finally, we display the state (f), which is localized near the ERM saddle points and constitutes a border between the S2 emergent states and the eigenstates, corresponding to the `background levels' discussed above, coexisting with them in energy and localized toward the outer edge of the Hamiltonian profile; An example of such state is the Wigner function (h) at approximately $e_{\rm vac}$.

%However, for a general $\delta\neq0$, $\lambda>\tilde{\lambda}_0$ and sufficiently large $\Delta$, we find specific Peres lattice points in the $(e_{\rm sad}, e_{\rm vac})$ interval with significantly lower values of $\langle \hat{n}\rangle$ than their neighbors in the spectrum; an approximate bound can be determined as $\langle\hat{n}\rangle\leq\langle\hat{n}\rangle_{\rm sad}=\Delta p_c^2/2$, shown in (b): These correspond precisely to the S2 emergent states. 

%

% and is accompanied by level bunching (finite-$\Delta$-precursor of the semiclassical log-singularity)

%the emergent states form the only subset of eigenstates that bear phase-space similarity to some of the $\delta=\pm1$ eigenstates (pure JC and aJC, respectively) at energies $e \leq e_\mathrm{vac}$.
%, giving them a clear interpretation: .   

%provides a cornerstone for possible experimental analysis, and  practical applicability of ESQPT-criticalities in the Extended Rabi Model. 

In Fig.~\ref{fig:5} (a) and (b), we also see points at $e>1/2$ and low $\langle\hat{n}\rangle$ attributed to the low-excitation eigenstates of the non-critical $m'=+1/2$ \eqref{7} subsystem, which appear regardless of the ESQPTs present in the critical case $m'=-1/2$ and are therefore of no use in the process of probing the criticalities. However, distinctive properties of the S2- and S2'-emergent states, differentiating them markedly from other eigenstates of the ERM Hamiltonian~\eqref{6} allow to construct experimentally relevant ESQPT-witness observables and are of significant consequence in critical driving protocols, as will be detailed in the following Sections. %In summary, the emergent states occur in a bounded energy interval between the two ESQPTs at $(e_{\rm sad}, e_{\rm vac})$ and for control parameter values $\lambda > \lambda_0$ and $|\delta| \in (0,1)$, i.e. between (but excluding) the pure Rabi and pure (anti) Jaynes-Cummings regimes. 
%{\color{blue} Significance of the emergent states in critical driving protocols will be demonstrated in the following section.}

\section{Witnesses of ESQPTs: Time-dependent protocols}\label{sec:5}

In order to effectively analyze the phase structure of the system under discussion, we must be able to recognize the dynamical signatures imprinted on a given state during its evolution through critical points as the interaction strength increases. At the same time, we aim to address the influence of the rate of increase of the interaction strength on the suitability of a given protocol for studying either ground-state QPT or the ESQPT criticalities. The problem of interest is therefore to solve the time-dependent Schrödinger equation
\begin{equation}\label{Schro}
i\frac{d}{d\tau}|\psi(\tau)\rangle=\hat{h}(\tau)|\psi(\tau)\rangle\,,\quad|\psi(0)\rangle=|\psi_0\rangle\,,
\end{equation}
\noindent where 
\begin{equation}\label{TimeConversion}
\tau=\varepsilon t\sqrt\Delta/\hbar    
\end{equation} is a rescaled (characteristic) evolution duration. The time dependence in the Hamiltonian \eqref{6} is assumed to enter through the interaction strength $\lambda(\tau)=\lambda_f\tau/\tau_f$, where $\tau_f$ is the interaction onset time, which determines its rate of increase. Since the temporal rescaling corresponds to multiplying by the qubit frequency in the rotated evolution frame, $\tau_f/(2\pi)$ represents the number of complete oscillations in this frame. Unless specified otherwise, in this paper, the initial state coincides with the non-interacting ground state \footnote{By choosing the non-interacting ground state, we do not imply that it is optimal for application of all the discussed critical phenomena in quantum state manipulation, however, it is definitely the best state for acquiring a proof of concept experimentally due to its enhanced dephasing resilience, e.g. compared to higher Fock states. Detailed optimization of the protocols and initial states will be given in a subsequent publication.}
\begin{equation}\label{psi0}
|\psi_0\rangle=|\downarrow\,,0\rangle\,.    
\end{equation}
The results presented in connection with time-dependent protocols can be understood in terms of the so-called strength functions (or local densities of states), which are defined as
\begin{equation}\label{StrF}
    P(e)=\sum_i |\langle\psi_i(\lambda_{f})|\psi(\tau_f)\rangle|^2\,\delta(e-e_i(\lambda_{f}))\:.
\end{equation}
\noindent These functions express the probability of finding the system state prepared at $\tau_f$ in the individual eigenstates of the Hamiltonian at the final interaction strength $\lambda_{f}$.

\begin{figure}[t!]
\begin{center}
    \includegraphics[width=\linewidth]{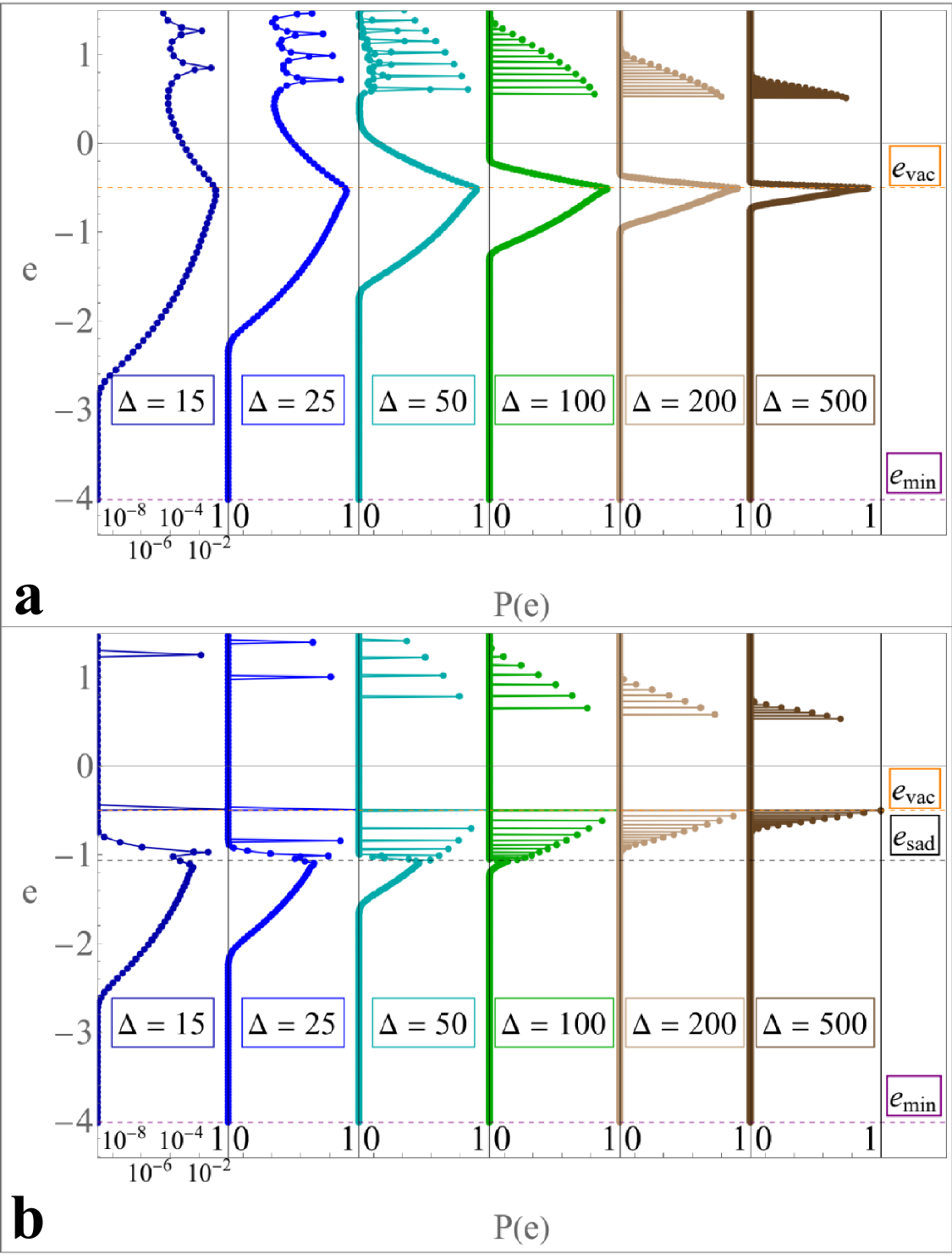}
%     \includegraphics[width=\linewidth]{Fig5_Fidelity.pdf}
%     \includegraphics[width=\linewidth]{Fig5_VacSurvival.pdf}
%  %\epsfig{file=F1a.eps,width=\linewidth}
%  %\epsfig{file=F1b.eps,width=\linewidth}
\end{center}
\caption{Strength functions, Eq.~\eqref{StrF}, in instantaneous quench protocols \mbox{($\tau_f=0$)} corresponding to a sudden increase of the interaction parameter to $\lambda_{f}=4$ at (a) $\delta=0$ and (b) $\delta=0.5$, for different system sizes $\Delta$. Horizontal dashed lines indicate the semiclassical critical points: local maximum [$e_{\rm vac}$ in (b)], saddle point [$e_{\rm sad}$ in (b) and $e_{\rm vac}$ in (a)], and global minimum ($e_{\rm min}$).} 
\label{fig:6}

\end{figure}

To demonstrate the significance of the S2 emergent states bound to semiclassical local maximum, at first we compare the strength functions corresponding to instantaneous ($\tau_f=0$) protocols in Fig.~\ref{fig:6}. For varying system sizes $\Delta$ and identical interaction parameters $\lambda_{f}=4$ and $\delta=0$ in panel (a), $\delta=0.5$ in (b), we observe the projections of $\psi_0$, Eq.~\eqref{psi0}, onto the eigenstates of the final Hamiltonian. The results in (b) confirm that stabilization of $\psi_0$ occurs in S2, and a set of emergent states with non-zero projection onto $\psi_0$ is found in the energy interval $(e_{\rm sad}, e_{\rm vac})$. Below $e_{\rm sad}$, in the region of S1-type excitations, the overlap is continuously suppressed with decreasing energy in a manner similar to the behavior of the strength function below $e_\mathrm{vac}$ in (a). As the system size $\Delta$ increases, the number of the S2 emergent states grows, and those with significant $\psi_0$ projections are pushed closer to $e_{\rm vac}$. In Fig.~\ref{fig:6}, we again see the non-critical states at $e>1/2$ that are localized near the origin of phase space, since they correspond to the excitations of the semiclassical Hamiltonian of the second subsystem with $m'=+1/2$ with just a global minimum in the origin. In protocols with finite system size and duration, these positive-energy states tend to decay into the $e<-1/2$ part of the spectrum, and the longest strength function trapping occurs in the aforementioned S2 emergent states (see below).

%\begin{itemize}
%    \item{}
%\end{itemize}
\begin{figure}[t!]
  \centering
  \includegraphics[width=\linewidth]{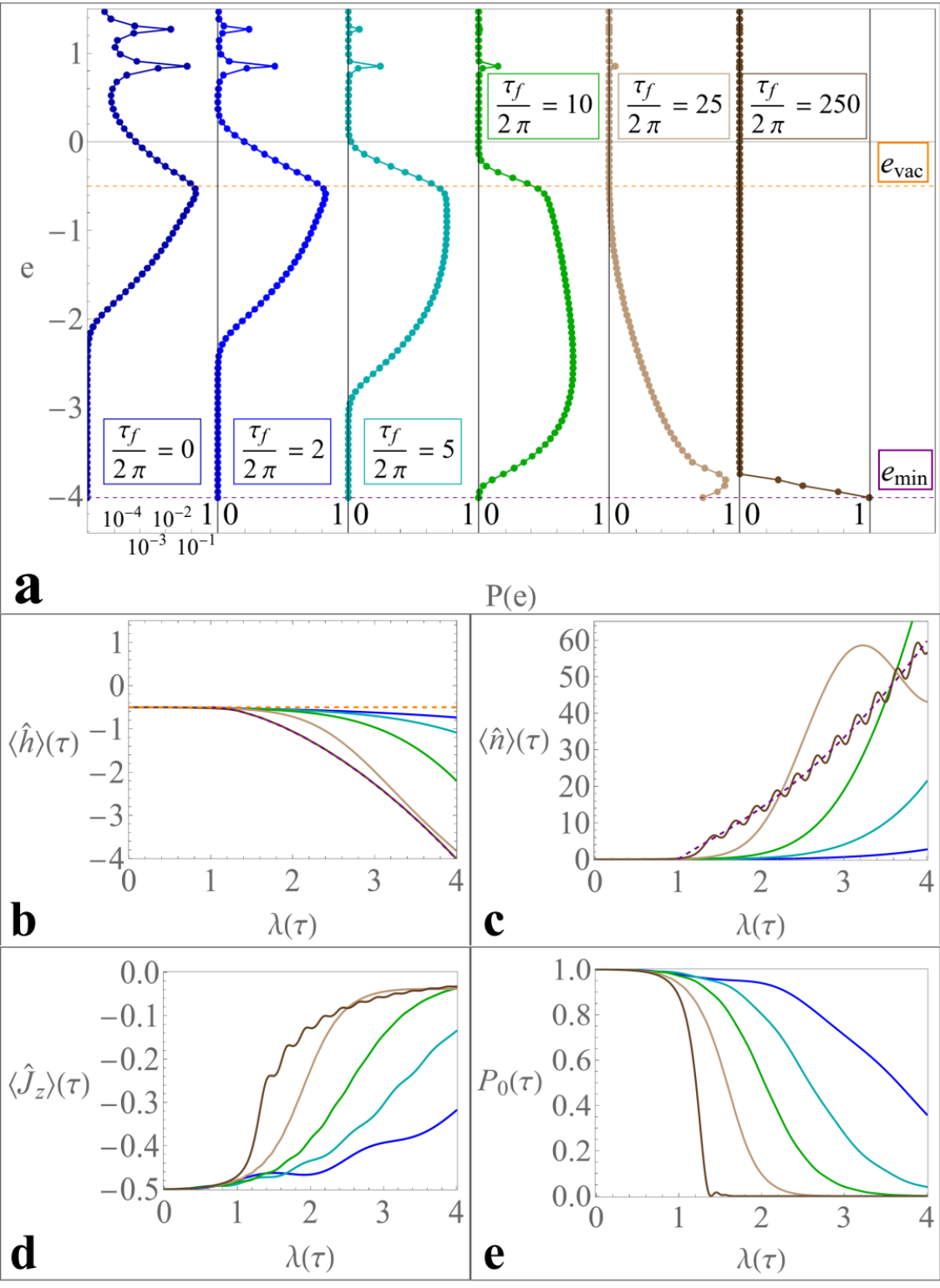}
  \caption{Strength functions, Eq.~\eqref{StrF}, resulting from linear-interaction-ramp driving protocols (a) with column-wise differing finite durations $\tau_f$ but with matching ERM parameters $\Delta=15$, $\lambda_f=4$ and $\delta=0$.  Panels in the lower half display $\lambda(\tau)$-parametrized evolutions of key quantities during the driving, namely the expectation values of (b) the time-dependent Hamiltonian, (c) the boson-number operator, (d) the quasispin operator and (e) the bosonic vacuum projector (or the $\psi_0$ survival probability). Orange and purple dashed lines in (a) and (b) are the semiclasically determined saddle-point and global minimum energies, respectively. Purple dashed line in (c) is the expected ground-state bosonic population.}
  \label{fig:7a}
\end{figure}

\begin{figure}[t!]
  \centering
  \includegraphics[width=\linewidth]{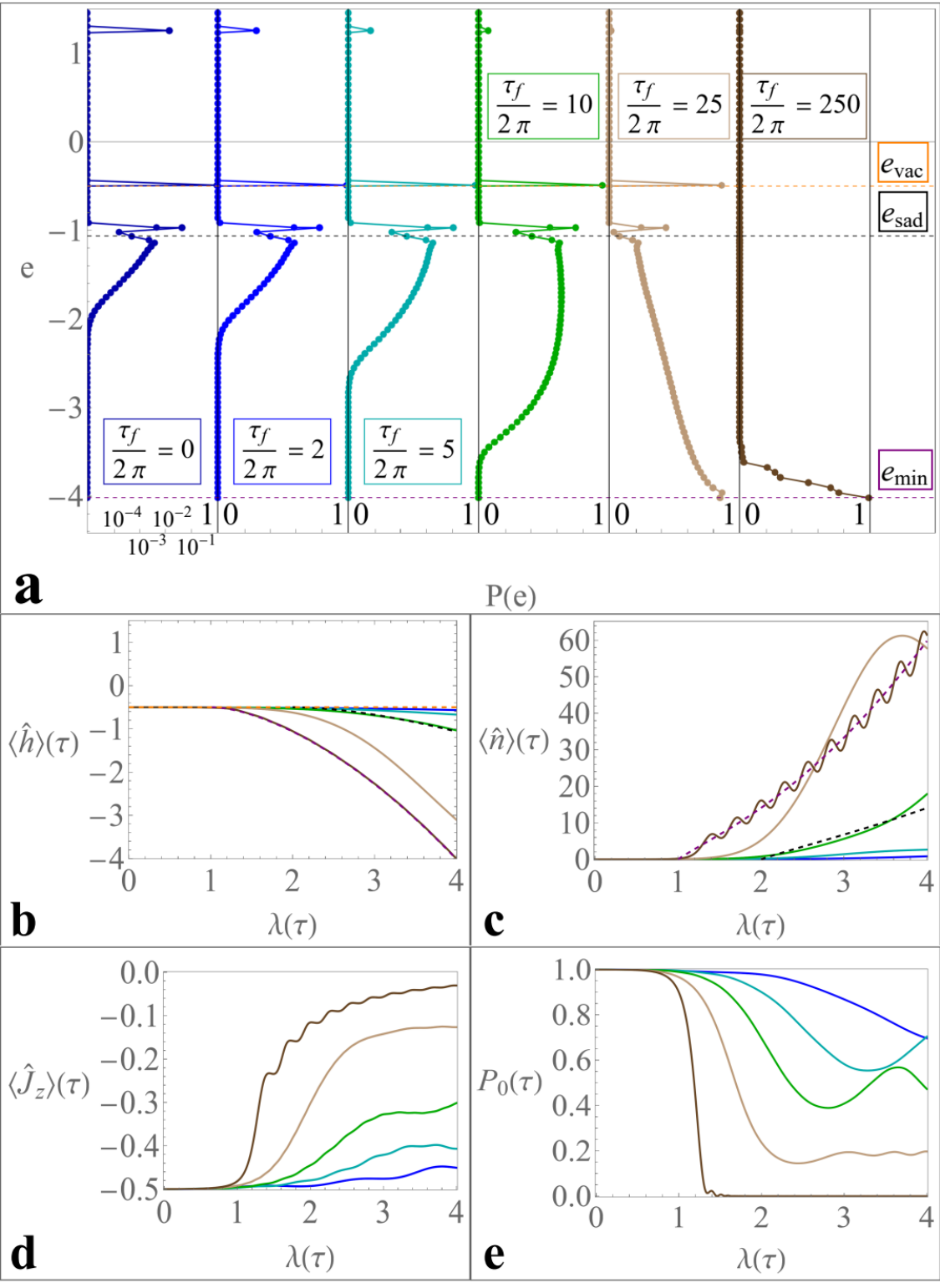}
  \caption{Same as in Fig. \ref{fig:7a} but with the $\delta=0.5$ interaction regime. In (a) and (b), the orange dashed lines therefore now represent the semiclassically determined local maximum energy and the black dashed lines the saddle point. The bosonic population assigned to the saddle-point phase-space coordinates is shown as the black dashed line in (c).}
  \label{fig:7b}
\end{figure}

In Figures \ref{fig:7a} and \ref{fig:7b}, we show the results of two sets of linear driving protocols with matching system sizes $\Delta=15$, target interaction strengths $\lambda_f=4$ and $\delta=0$ in Fig. \ref{fig:7a} and $\delta=0.5$ in Fig. \ref{fig:7b}. Strength functions are shown in both figures for 6 different values of protocol duration $\tau_f$ ranging from instantaneous quenches with $\tau_f = 0$ (leftmost in panels a) to $\tau_f^{(\rm max)}=500\pi$ (rightmost in panels a). The slowest protocols correspond to near-adiabatic drive for both the $\delta = 0$ and $\delta = 0.5$ cases shown. The significance of the cascade of emergent states in the interval $(e_{\rm sad}, e_{\rm vac})$ lies in their ability to dynamically capture the wave function during driving from $\psi_0$ of Eq.~\eqref{psi0}. The strength functions corresponding to the standard Rabi model steadily proceed toward the ground state as $\tau_f$ increases. In contrast, in the extended Rabi model, where $e_{\rm vac}$ is a semiclassical maximum, trapping between ESQPT energies occurs and naturally hinders adiabatic behavior. To achieve truly adiabatic evolution in such a driving regime, the increase of interaction strength - especially in the $(\tilde{\lambda}_c,\,\tilde{\lambda}_0)$ interval - must be sufficiently slow, allowing the wave function to leave the $e_{\rm vac}$ region and avoid the subsequently stabilized vacuum and the other S2 emergent states. Driving protocols with a sufficiently fast transition to S2 are therefore particularly suitable for probing the spectral region between ESQPTs, as the wave function is naturally trapped there. We note that the rate of decay of the non-critical $m'=+1/2$ states at $e>1/2$ is significantly larger than that of the S2 emergent states as evident by the log-scale in \mbox{Fig. \ref{fig:7b} (a).} Their decay to the $e<-1/2$ part of the spectrum is also only weakly affected (accelerated) by the presence of the emergent states, as can be seen by comparison with \mbox{Fig. \ref{fig:7a} (a).}

% \begin{figure*}[h!]
%   \centering
%   \includegraphics[width=0.49\linewidth]{Fig7b.pdf}
%   \includegraphics[width=0.49\linewidth]{Fig7a.pdf}
%   \caption{...}
%   \label{fig:7x}
% \end{figure*}

The lower halves of Figures \ref{fig:7a} and \ref{fig:7b} show the time-dependent expectation values of the instantaneous Hamiltonian $\langle \hat{h} \rangle(\tau)$, the phonon number $\langle \hat{n} \rangle(\tau)$, the qubit orientation $\langle \hat{J}_z \rangle(\tau)$, and the projection $P_0(\tau)$ onto $\psi_0$ (survival probability) for each set of protocols in panels (b)-(e), respectively. In the $\delta=0.5$ case, there exists a theoretical (critical) duration $\tilde{\tau}_f$, for which both the $\langle\hat{h}\rangle(\tau)$ and $\langle\hat{n}\rangle(\tau)$ curves approximately follow the characteristics of the saddle point in S2, thus the lower ESQPT critical energy $e_\mathrm{sad}$. Protocols in which the interaction strength increases at a rate slower than this supposed threshold are not well suited to probe the ESQPT region, but are feasible for observing ground-state QPT effects~\cite{Duan21}. Based on the above observations, we infer that the main quantities to serve as ESQPT witnesses are the mean phonon population $\langle\hat{n}\rangle$ and the vacuum survival probability $P_0$. For $\tau_f<\tilde{\tau}_f$, we observe only a moderate increase in mean phonon numbers $\langle\hat{n}\rangle$ in protocols ending within S2/S2' phases (cf. Figs. \ref{fig:7a} and \ref{fig:7b}, panels c). Furthermore, among the $\delta=0.5$ protocols shown in Fig. \ref{fig:7b}, only the slowest protocol with $\tau_f=500\pi$ allows the state to adapt fully to S1, so that the vacuum contribution is erased; in all other protocols, a nonzero projection onto the vacuum state $P_0$ persists in S2, in contrast to the standard Rabi protocols, where the vacuum eventually always tends to decay (cf. Figs. \ref{fig:7a} and \ref{fig:7b}, panels e).

\section{Suitable experimental observables}\label{sec:ExptParamMaster}
In this section, we discuss influences due to realistic decoherence and dephasing typical of the current state-of-the-art experimental setups %within the Monte Carlo wave function (MCWF) approximation of the Lindblad master equation introduced by M{\o}lmer, Castin and Dalibard~\cite{Molmer}
for the extended Rabi model in Paul trap setups similar to Innsbruck/Brno trap~\cite{Obsil} (using $^{40}$Ca$^+$) and the Tsinghua trap~\cite{Duan21} (using $^{171}$Yb$^+$). We conclude the section by proposing experimentally verifiable dependencies of ESQPT witnesses.

The inevitable coupling of the trapped ion to the environment implies that, in order to make realistic predictions about experimental outcomes, we must prescribe a suitable master equation to describe the evolution of the open quantum system. Since a trapped ion is generally weakly coupled to its environment~\cite{GardinerZollerBooks}, which can be considered Markovian (memoryless), we employ the Lindblad master equation in the Schrödinger picture
\begin{equation}\label{15}
    \frac{d}{dt}\hat{\rho}_s(t)=-\frac{i}{\hbar}\left[\hat{H}_s(t),\hat{\rho}_s(t)\right]+\mathcal{L}\left[\hat{\rho}_s(t)\right]\,,
\end{equation}
\noindent where the Lindblad superoperator $\mathcal{L}$ takes the form
\begin{equation}\label{dissip}
    \mathcal{L}\left[\hat{\rho}_s(t)\right]=\sum_j\left(\hat{L}_j\hat{\rho}_s(t)\hat{L}_j^\dagger-\frac{1}{2}\left\{\hat{L}_j^\dagger\hat{L}_j,\hat{\rho}_s(t)\right\}\right)\,,
\end{equation}
with $\hat{L}_j$ representing individual dissipators acting on the density matrix of the system $\hat{\rho}_s$ and $\{\hat{A},\hat{B}\}=\hat{A}\hat{B}+\hat{B}\hat{A}$ denoting the anticommutator.

Environmental effects that should be taken into account in a comprehensive error analysis include~\cite{GardinerZollerBooks}:
\begin{itemize}
    \item \textit{Motional dephasing}: $\hat{L}_m=\sqrt{2\Gamma_m}\,\hat{a}^\dagger\hat{a}$, arising from (trap) secular-frequency noise.
    \item \textit{Heating and damping}: \mbox{$\hat{L}_h=\sqrt{\gamma n_{th}}\,\hat{a}^\dagger$,} \mbox{$\hat{L}_d=\sqrt{\gamma (n_{th}+1)}\,\hat{a}$,} which result from incoherent energy exchange with the environment. This form models the environment as a large thermal bath with a population $n_{th}$, coupled to the system at a characteristic rate $\gamma$.
    \item \textit{Qubit dephasing}: $\hat{L}_q=\sqrt{2\Gamma_q}\,(\hat{J}_z+1/2)$, stemming from laser-field and magnetic-field fluctuations affecting the ion's internal states.
\end{itemize}
\noindent Furthermore, we note that the omission of the qubit-relaxation dissipator $\hat{L}_r\propto\hat{J}_-$ is justified by the fact that the qubit excited-state lifetime in a trapped ion is expected to be much longer than the characteristic experimental timescales (unlike, for example, in superconducting circuits \cite{Cernotik}).

Equation \eqref{15} can be transformed in accordance with the evolution-frame rotations performed in Sec. II., since these modify the prescribed dissipators only by a phase factor. As a result, we can define $\hat{l}_j=\left(\frac{\varepsilon\sqrt{\Delta}}{\hbar}\right)^{-1/2}\hat{L}_j$, which leads to
\begin{equation}\label{Lindblad}
    \frac{d}{d\tau}\hat{\rho}(\tau)=-i\hat{\mathfrak{h}}(\tau)\hat{\rho}(\tau)+i\hat{\rho}(\tau)\mathfrak{\hat{h}}^\dagger(\tau)+\sum_j\hat{l}_j\hat{\rho}(\tau)\hat{l}_j^\dagger\,,
\end{equation}
\noindent where $\hat{\mathfrak{h}}(\tau)=\hat{h}(\tau)-\frac{i}{2}\sum_j\hat{l}_j^\dagger\hat{l}_j$ is an effective non-Hermitian Hamiltonian. Separating the master equation into non-Hermitian Hamiltonian propagation and \textit{quantum jump} term motivates the Monte Carlo wave-function method~\cite{Molmer} for its solution, which we discuss in more detail in the Appendix~\ref{App:B}.

Making predictions about a realistic linear-interaction-ramp protocol requires us to determine sideband-drive detunings $\delta_{r,b}$ and maximal Rabi frequencies $\eta\Omega_{r,b}^{(\rm max)}$ as well as the ramp duration $t_f$ and the above mentioned characteristic dephasing and heating rates. Since the rates of environmentally caused motional decoherences scale at least as $\sqrt{n}$ (and up to $n^2$) for the $n$th-Fock components in $\hat{\rho}$, obtaining precise phonon-basis statistics can be challenging. Therefore, we propose the projection of the final state on the vacuum to be the main experimentally observable witness of ESQPT.
%the survival probability of the vacuum state to be the main experimental observable of interest. (změkčit)
\begin{figure}[b!]
  \centering
  \includegraphics[width=\linewidth]{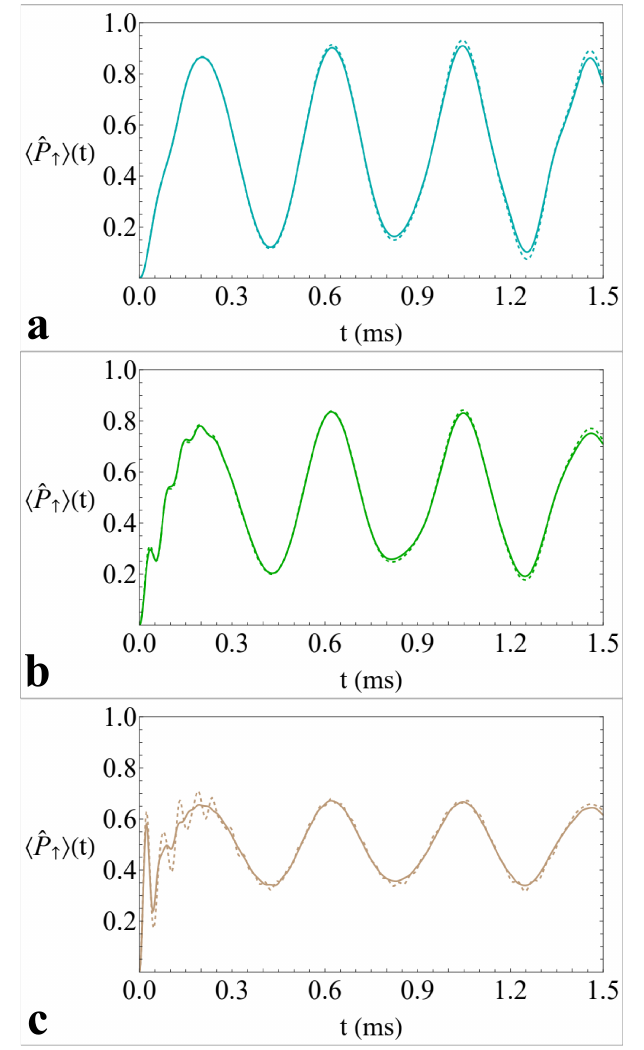}
  \caption{Rabi oscillations numerically obtained by evolving a blue-sideband interaction at fixed Rabi frequency \mbox{$\eta\Omega_2=2\pi\times2.4\:\rm kHz$.} Initial states are the qubit-down projected results of linear-interaction-ramp protocols simulated for matching experimental parameters \mbox{$\delta_r=-2\pi\times3.6\:\rm kHz$,} $\delta_b=-2\pi\times4.1\:\rm kHz$, $\eta\Omega_1^{(\rm max)}=2\pi\times5.87\:\rm kHz$ and \mbox{$\eta\Omega_2^{(\rm max)}=2\pi\times1.96\:\rm kHz$,} but for different Rabi frequency onset durations (a) $t_f=1.3\: \rm ms$, (b) $t_f=2.6\: \rm ms$, (c) $t_f=6.5\: \rm ms$. Corresponding ERM parameters are $\varepsilon/\hbar=2\pi\times0.98\:\rm kHz$, $\Delta=15.4$, $\lambda^{(\rm max)}=4.0$ and $\delta=0.5$, along with the rescaled durations (a) $\tau_f/(2\pi)=5$, (b) $\tau_f/(2\pi)=10$, (c) $\tau_f/(2\pi)=25$. Dashed curves correspond to Schrödinger-equation evolution \eqref{Schro} both during the ERM protocol and the diagnostic blue-sideband drive. Full curves display corrections stemming from the Lindblad master equation \eqref{Lindblad} again both during the protocol and the diagnostic drive. Considered environmental effects and their estimated rates include motional dephasing with $\Gamma_m^{-1}=100\:\rm ms$, qubit dephasing with $\Gamma_q^{-1}=10\:\rm ms$ and motional heating rate $\gamma n_{th}=3.3\:\rm s^{-1}$.}
  \label{fig:8}
\end{figure}

Experimental post-driving state diagnostics that yield phonon statistics can be performed by measuring blue-sideband-driven Rabi oscillations and fitting the obtained curve \cite{Duan21,dingshun}. To simplify the data-fitting procedure, before the diagnostic drive, it is customary to project the state of the system on the qubit-down state, all the while preserving the motional degree of freedom unaffected. Note, however, that this projection, although crucially simplifying the process of obtaining phonon-basis statistics, non-trivially alters the state and e.g. erases qubit-motion entanglement. In context of the Hamiltonians we discussed in Sec. II., the blue-sideband drive corresponds to \eqref{3} with $\delta_{r,b}=0$ and $\Omega_1=0$. Each individual $n$-phonon component adds to the Rabi oscillations through a $\eta\Omega_2\sqrt{n+1}\:$-frequency contribution with amplitude equal to the population $p_n$. Extracting the vacuum population, in particular, requires only the 'contrast' of such Rabi oscillations at longer driving times. The vacuum contribution possesses the longest-coherence-time component, due to its natural dephasing resilience.

For a set of model-suitable and experimentally feasible parameters, we show in Fig. \ref{fig:8} the post-interaction-ramp Rabi oscillations at fixed blue-sideband Rabi frequency for three different protocol durations $\tau_f/2\pi = 5,10,25$ in panels (a-c), respectively, with prediagnostic qubit-down projection. The Appendix~\ref{App:C} further elaborates upon our specific choice of parameters and their possible adjustments with respect to preserving the same ERM Hamiltonian through \eqref{eq:Lambdadelta}-\eqref{eq:lam}. Comparison of results obtained by solving the Schrödinger equation (dashed lines) and the Lindblad master equation (solid lines) shows that, for a state-of-the-art trapped ion setup, environmental effects can be neglected when only the vacuum contribution is extracted. Moreover, the qubit-down projection before the diagnostic blue-sideband drive removes the initial phase mismatch, acquired due to environmental effects during the critical-driving protocol. The remaining non-unitary effects are then the overall damping and decoherence of higher Fock components, especially pronounced for longer protocols (panel c), neither of which impacts significantly the total contrast (width of $\langle \hat{J}_z \rangle(t)$ oscillations) determining the sought after vacuum contribution.

Generalization of the results presented in Figures \ref{fig:7a} (e) and \ref{fig:7b} (e) to different $\tau_f, \delta$ and $\Delta$ provides theoretical predictions for a feasible experimental verification of the ESQPT signatures in the ERM. For the initial state $\psi_0$ of Eq. \eqref{psi0} undergoing a linear-interaction-ramp protocol reaching a fixed $\lambda_f$ in an ERM defined by system size $\Delta$ and interaction regime $\delta$, the $\tau_f$-dependence of the vacuum population is uniquely determined by the encountered critical points. In accordance with the post-driving measurement, we define the down-projected vacuum population $\tilde{P}_0(\tau_f)=P_0(\tau_f)/\langle\hat{P}_\downarrow(\tau_f)\rangle$, where $P_0(\tau_f)$ is the survival probability at the end of the interaction ramp. The division by qubit-down population results from the prediagnostic projection.

\begin{figure}[t!]
  \centering
  \includegraphics[width=\linewidth]{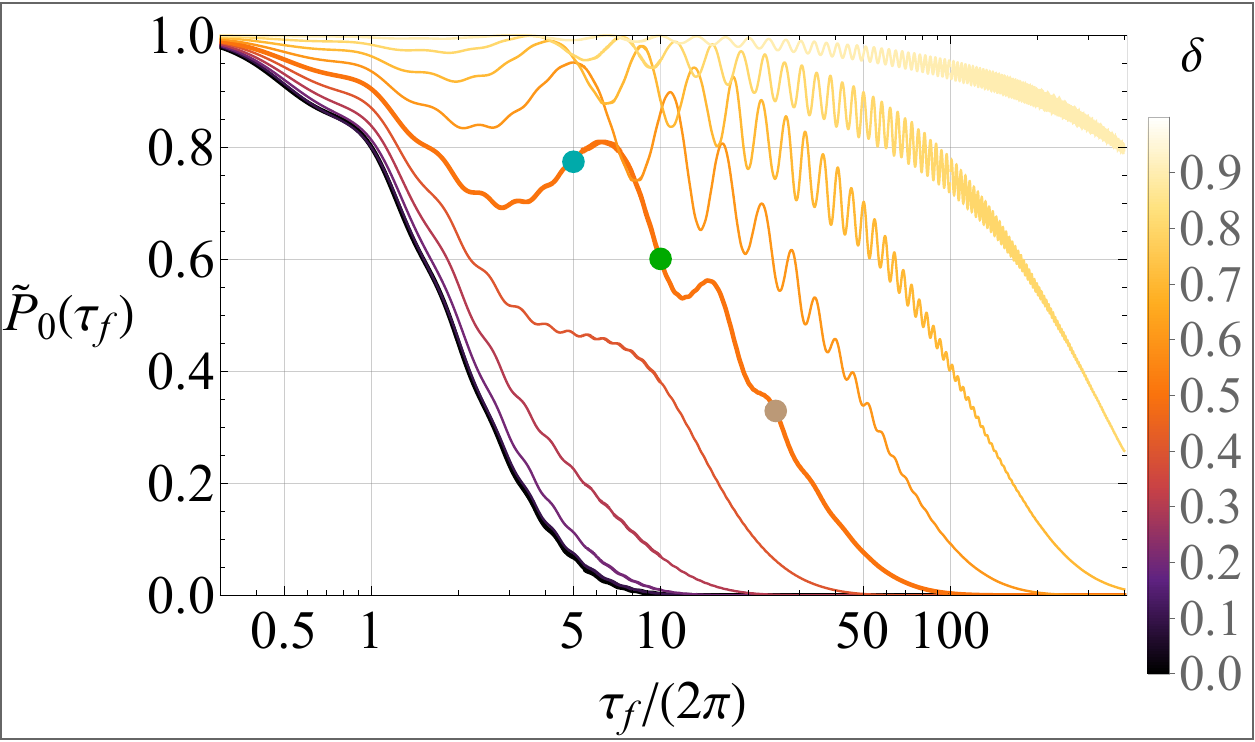}
  \caption{Vacuum populations of the down-projected states resulting from linear-interaction-ramp protocols \eqref{Schro} depending on protocol duration $\tau_f$. States $|\psi(\tau_f)\rangle$ are obtained for $\Delta=15.4$ and final interaction strength $\lambda_f=4$, but for various $\delta\geq0$, which color-code the individual curves. Regimes $\delta\in\{0,\,0.5\}$ are highlighted by line thickness and points corresponding to Rabi oscillations in Fig. 5 are marked.}
  \label{fig:9}
\end{figure}

Figure \ref{fig:9} shows $\tilde{P}_0(\tau_f)$ resulting from solutions of \eqref{Schro} for different $\delta\geq0$ interaction regimes and fixed $\Delta=15.4$, $\lambda_f=4$. Since the semiclassically determined ground-state energy and phase-space position ($x=\pm x_c$, $p=0$) are independent of $\delta$ and $|\langle\psi_{gs}|\psi_0\rangle|^2\approx0$ at $\lambda=4$ (see Fig. \ref{fig:6}), the decay of $\tilde{P}_0$ is a necessary condition for high-fidelity adiabatic evolution regardless of the specific $\delta$. The immediate consequence of Fig. \ref{fig:9} is then the demonstration of the drastic increase with $\delta$ in the lower-threshold duration below which the adiabatic behavior is forbidden by the ESQPT structure. More generally, the variation of $\tilde{P}_0(\tau_f)$ with $\delta$ becomes abrupt when we cross the finite-size corrected S2 boundary. As a result of the wave function being trapped by the emergent states, we see plateaus and local maxima in $\tilde{P}_0(\tau_f)$ for $\delta\gtrsim \lambda_f^{-1}$, the limit of which is $\tilde{P}_0(\tau_f)=1$ for $\delta=1$ as $\psi_0$ \eqref{psi0} is uncoupled by the pure Jaynes-Cummings interaction.

\begin{figure}[h!]
  \centering
  \includegraphics[width=\linewidth]{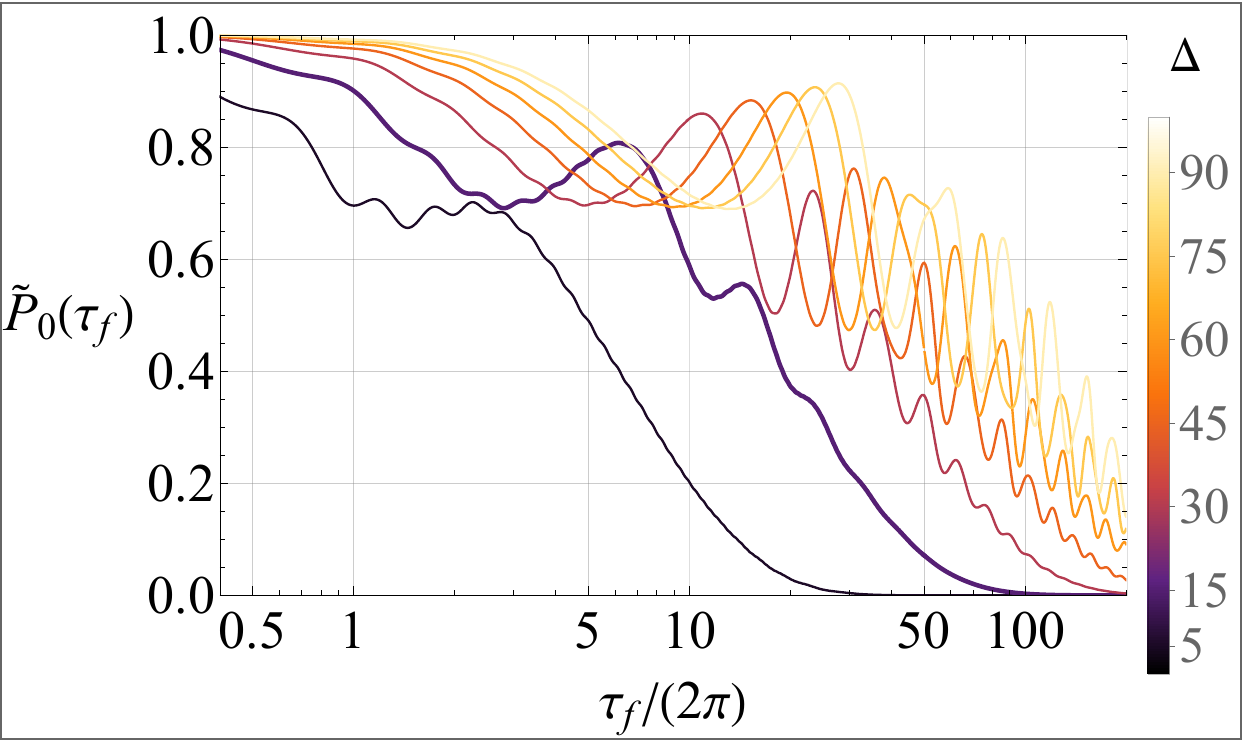}
  \caption{Same as in Fig. \ref{fig:9}, but for fixed $\delta=0.5$ for various system sizes $5\leq\Delta\leq90$. Regime $\Delta=15$ is highlighted by line thickness.}
  \label{fig:10}
\end{figure}

Finite-size scaling behavior of the rich ESQPT structure in the ERM can be studied from the same dependencies for a fixed interaction regime and varying system size. In Fig. \ref{fig:10}, we show the $\tilde{P}_0(\tau_f)$ curves for fixed $\lambda_f=4$, $\delta=0.5$ but for different $\Delta$. Although the limiting classical evolution $\Delta\to\infty$ is again $\tilde{P}_0(\tau_f)=1$, since $h(0,0,m')$ in \eqref{7} represents a stationary point regardless of $\lambda(\tau)$ or $\delta$, we see in Fig. \ref{fig:10} that the finite width of the Wigner function in the examined $\Delta$-range eventually still leads to the $\tilde{P}_0(\tau_f)$ decay. However, a key aspect of the finite-size scaling of $\tilde{P}_0(\tau_f)$ is the increasing height of the peaks associated with the trapping of the wave function and the general shift toward larger values of $\tau_f$ in accordance with the classical limit.

\section{Conclusions}\label{sec:Concl}
%\begin{itemize}
%    \item{g.s. QPT needs to be probed by slow driving}
%    \item{ESQPT needs to be probed by fast driving (quenches)}
%\end{itemize}

We have shown that the externally driven Extended Rabi Hamiltonian, using protocols linear in time and with adjustable speed, enables experimental realization and study of ESQPT criticalities accompanying the ground state QPT between the normal and superradiant phases. In the control parameter region corresponding to the superradiant phases of the model Hamiltonian,  the excited superradiant phase of the (anti) Jaynes-Cummings type S2 (S2') occurs energy-wise above the Rabi superradiant phase S1, and the transition between these phases can be detected by addressing several ESQPT-witness observables: (i) the vacuum survival probability, (ii) expectation value of the phonon number operator, (iii) the quasispin projection operator. All of them can be extracted from experimentally recorded Rabi oscillations. 

On top of the vacuum stabilization under sudden quenches identified in~\cite{Stra21}, the key new contribution of this paper is the identification of a family of `S2-emergent states', which are addressed by driving protocols at finite speed. The emergent states share a non-zero projection onto the vacuum $\ket{\downarrow\,,0}$ itself, however their occurrence is not limited to the energy of the stationary point of the Hamiltonian at the origin, but spans the whole energy range precisely between the two critical ESQPT energies $e_\mathrm{vac}$ and $e_\mathrm{sad}$. In the phase space, their Wigner functions are concentrated around the local maximum of the semiclassical Hamiltonian at the origin (at the upper ESQPT energy $e_\mathrm{vac}$) up to the location of the saddle points (lower ESQPT energy $e_\mathrm{sad}$) and squeezed in the $x-$direction. %, distinguishing them sharply from other eigenstates of the model coexisting with them energy- and parameter-wise
Thus, by carefully adjusting the driving protocol speed, population of the S2 (S2') emergent states allow to probe both ESQPT criticalities. In other words, they represent a fingerprint for entering the excited superradiant (a)JC phases S2 and S2'. Considering the typical properties of current state-of-the-art ion trap setups, we have shown that the non-Hermitian effects (dissipation, dephasing) contribute only weakly to the overall signatures of the criticality. 

We have studied here the particular and model specific signatures of ESQPT quantum criticality in the simplest possible setting of a single cold trapped ion, realizing the ERM Hamiltonian. We have shown that the plentitude and robustness of the S2 emergent states of the ERM allows to experimentally realize ESQPTs, similar to ground state QPTs with the trapped ion setup~\cite{Duan21}. This opens up a particularly interesting possibility of high-fidelity preparation of a `near-saddle-point' emergent state, which would provide an improvement in displacement sensing applications with respect to typically used non-Gaussian (multiple Fock) states~\cite{FilipManyFock}, especially in the context of the developing field of critical quantum sensing~\cite{Gietka25}. 
In more general terms, we believe that the conceptual simplicity of the ESQPT approach (based essentially on determining the unstable equilibria of the semiclassical limit in the control parameter and excitation energy spaces) to quantum criticalities provides an intuitive roadmap for quantum state preparation and manipulation with interesting opportunities in diverse quantum technology, metrology, and sensing applications based on quantum critical phenomena~\cite{Frerot18,Beaulieu25,Gietka25,FilipManyFock} in particular with cold trapped ions, but also other setups, e.g. using the superconducting circuits~\cite{Chavez-Kerr, Iachello-Kerr-1, Iachello-Kerr-2}.   

\section{Acknowledgments}

We thank P. Cejnar, P. Str\'ansk\'y, L. Slodi{\v c}ka, V. {\v S}varc, T. M. Pham, O. {\v C}\'ip, A. {\v C}epil, O. {\v C}ernot\'ik, and R. Filip for insightful discussions and comments. This work was supported by the Czech Ministry of Education, Youth and Sports, project No. CZ.02.01.01/00/22 008/0004649 (QUEENTEC).
Authors also acknowledge the project 23FUN03 HIOC, which has received funding from the European Partnership on Metrology, co-financed from the European Union’s Horizon Europe Research and Innovation Programme and by the Participating States.
M.K. acknowledges partial support by the Czech Science Foundation project No. 25-16056S. 

\appendix 
\section{Characteristics of the Excited Superradiant (anti) Jaynes-Cummings phases S2 and S2'}\label{App:A}

In this section, we further address the presence and nature of the S2 (and S2') emergent states of \eqref{6}. From the phase-space viewpoint, their Wigner functions are largely trapped by the semiclassical-Hamiltonian separatrix in the region of the local maximum. Thus far, we have focused on the fact that the result of this trapping is a significantly lower bosonic population compared to their spectral neighbors. This energetic proximity of states with very different spatial characteristics can be understood analytically from the Jaynes-Cummings and anti-Jaynes-Cummings models, where the critical semiclassical Hamiltonian assumes the Mexican-hat shape (ERM saddle points are lost to the axially symmetric minimum).

The Hamiltonian \eqref{6} restricted to $|\delta|=1$ couples only states within two-dimensional subspaces of the original Hilbert space and can therefore be analytically solved. The resulting spectrum is
\begin{equation}\label{A1}
    \sigma_{\rm sign (\delta)}=\left\{-\frac{\rm sign (\delta)}{2}\right\}\:\cup\left\{\frac{2n+1}{2\Delta}\pm\frac{r_n}{2}\right\}_{n=0}^\infty\;,
\end{equation}
\noindent where $r_n=\sqrt{\left(\frac{1-\rm sign (\delta) \Delta}{\Delta}\right)^2+4\lambda^2\frac{n+1}{\Delta}}\;.$ The eigenvalue $-\rm sign(\delta)/2$ is derived from the fact that the qubit-down vacuum and the qubit-up vacuum are uncoupled for $\delta=1$ and $\delta=-1$, respectively.

The mean bosonic populations in the $\delta=\pm1$ eigenstates corresponding to \eqref{A1} are convex combinations of $n$ and $n+1$. The connection we want to emphasize is that in S2/S2', for every $e\in(e_{\rm min}, e_{\rm vac})$, we can find two separate circles described by $h(\sqrt{x^2+p^2})=e$ in the classical phase space. One of these circles will be closer (at every point) to the origin than the $e_{\rm min}$-circle, and the other will be farther away. Similarly, for $e(n_{\rm min})=\min(\sigma_{\rm sign (\delta)})$ we can find two real numbers $\tilde{n}_{1,2}\geq0$ by inverting \eqref{A1}, which satisfy $\tilde{n}_1\leq n_{\rm min} \leq \tilde{n}_2$ and at the same time $e(\tilde{n}_i)>e(n_{\rm min})$ for both $i=1,\,2$. In the $\Delta\gg1$ case there are actual eigenvalues $e_{n_1}$ and $e_{n_2}$ arbitrarily close to any $e_n$ satisfying $e_n\in(e_{\rm min}, e_{\rm vac})$, and $\langle\hat{n}\rangle_{\psi_{n_1}}\leq\langle\hat{n}\rangle_{\psi_{n_{\rm min}}}\leq\langle\hat{n}\rangle_{\psi_{n_2}}$. The eigenstates with $\langle\hat{n}\rangle$ lower than the ground state are the emergent states in the $\delta=\pm1$ regimes.

This intuition behind emergent states is slightly altered when we reintroduce the saddle points present at $(x',p')=(0,\pm p_c)$ in $\delta\notin\{0,\pm1\}$ cases. The heuristic bound for distinguishing emergent states in the $(e_{\rm sad},e_{\rm vac})$ interval comes from their phase-space confinement within the region of the local maximum and can be established for example as $\langle \hat{n}\rangle_\psi\lesssim \Delta p_c^2/2$, which is demonstrated in Fig. \ref{fig:5} (b). Another option would be to quantify the states' variations in $x'$ and $p'$ and limit it to a fraction of the width of the separatrix in the respective directions. The classical phase-space analogy can be utilized to approximate the number of emergent states at a given $(\lambda,\, \delta)$. We recall the $m'=-1/2$ semiclassical Hamiltonian \eqref{7} in S2/S2' \mbox{($\lambda>1/|\delta|$).} By transitioning to polar coordinates $(x',p')\to(r\cos\varphi,r\sin\varphi)$, we find that for each $e\in (e_{\rm sad},e_{\rm vac})$ there exist two solutions $r^\pm(\varphi,e)$ to $h(r,\varphi)=e$. Each of these solutions prescribes a closed curve in the region of phase space bounded by the separatrix ($r^+$ outward and $r^-$ inward), which merges with the separatrix when $e\to e_{\rm sad}$. Therefore, we can define the inner and outer phase-space volumes
\begin{equation}\label{A2}
    v^\pm (\lambda,\delta)=\int_{e_{\rm sad}}^{e_{\rm vac}}de\int_0^{2\pi}d\varphi f^\pm(\varphi,e)\,,
\end{equation}
\noindent where $f^\pm=r^\pm\left|\frac{\partial h}{\partial r}\right(r^\pm,\varphi)|^{-1}$. Integration in \eqref{A2} can be carried out exactly, and it is possible to roughly estimate the number of emergent states $v^-\Delta/(2\pi)$ and also its ratio to the number of 'background' states $v^-/v^+$ between ESQPT energies even for moderate $\Delta$. 

However, note that for systems outside the $\Delta\gg 1$ limit, the symmetry in $\pm\delta$ is broken. In particular, the $\delta<0$ spectra show primary stabilization of the qubit-up vacuum even for $\lambda<\lambda_c$ as opposed to the $\delta>0$ cases that we focus on in the article. Emergent states still occur when $\delta<0$, but we will observe one at the $e_{\rm vac}$ energy and of the qubit-down vacuum type only as the result of $\Delta$-asymptotics and not the direct stabilization beyond finite-size corrected $\tilde{\lambda}_0$ as we saw in Figs \ref{fig:2} (b), \ref{fig:5} (b), \ref{fig:5} (c), \ref{fig:6} (b). This consequence of broken $\pm\delta$ symmetry needs to be understood when we predict the number of emergent states for $\Delta\lesssim10^2$ from the classical approaches presented. On the other hand, $\delta<0$ finite-duration interaction ramps to S2' provide a unique possibility of avoiding the S2 vacuum stabilization at $e_{\rm vac}$ and accessing the next (wider) emergent state with high fidelity. 

\begin{figure}[t!]
  \centering
  \includegraphics[width=\linewidth]{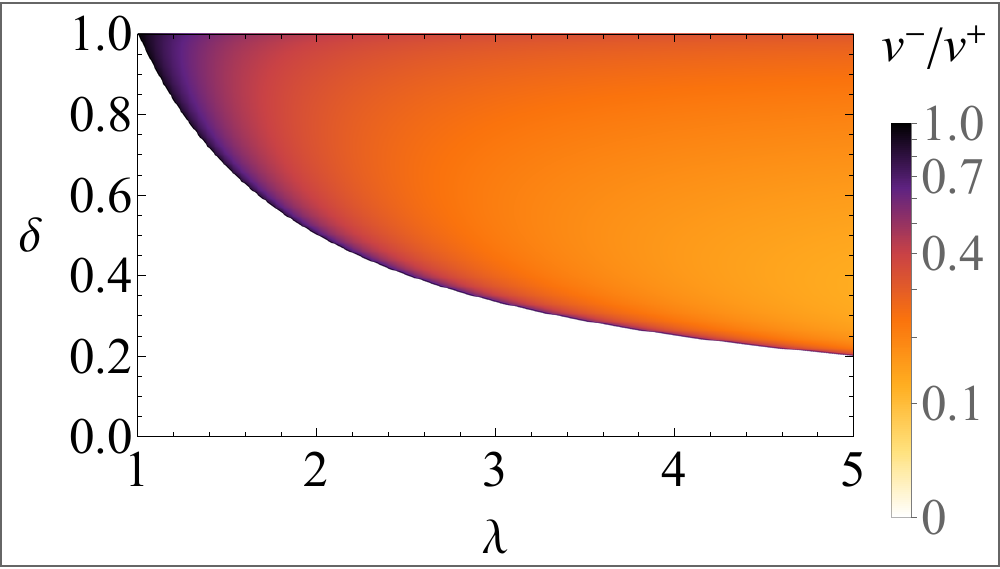}
  \caption{Color-coded map of the inner and outer phase-space volumes ratio $(v^-/v^+)(\lambda,\,\delta)$ defined through \eqref{A2}. The definition is valid only for $\lambda>1/|\delta|$.}
  \label{fig:app}
\end{figure}

We conclude this section by showing the $(\lambda,\delta)$-dependence of the ratio $v^-/v^+$, defined in \eqref{A2}, as a 2D colored map in Fig. \ref{fig:app}. It can be proved that $v^-=v^+$ in the limit $\lambda\to(1/\delta)^+$, but it is apparent that the share of emergent states quickly drops, when one moves above the $\delta=1/\lambda$ curve, which is required for finite $\Delta$ observations. Results based on phase-space volumes that predict the number of emergent states as $N_e/\Delta=v^-/(2\pi)$ and the number of 'background' states as $N_n/\Delta=v^+/(2\pi)$ were tested. From numerical diagonalization of \eqref{6} for $\lambda=4$ and $\delta=0.5$ in the range $\Delta\in[20\,,800]$, we validated the predictions of \eqref{A2}, $v^-(4,\,0.5)\doteq2\pi\cdot0.210$ and $v^+(4,\,0.5)\doteq2\pi\cdot1.335\,$. The eigenstates were classified according to the $\langle\hat{n}\rangle_\psi/\Delta\gtrless p_c^2/2$ criterion, and the resulting relative errors of the classical phase-space predictions are approximately $4.8\,\%$ for $\Delta=20$ and $0.6\,\%$ for $\Delta=800$. 

\section{Solving the Lindblad master equation}\label{App:B}

Here, we further elaborate upon the Monte Carlo wave function (MCWF) method utilized for solving \eqref{Lindblad} and first introduced in \cite{Molmer}. The crucial advantage of applying this method, is that one works only with individual wave functions, which is preferable to the entire density matrix for large systems in terms of computational memory. Eq. \eqref{Lindblad} admits a formal solution in the form of a series \mbox{$\hat{\rho}(\tau)=\sum_{n=0}^\infty\hat{\rho}^{(n)}(\tau)$.} If we define $\hat{u}(\tau,\tau_0)$ as the solution to \mbox{$\partial_\tau\hat{u}(\tau,\tau_0)=-i\hat{\mathfrak{h}}(\tau)\hat{u}(\tau,\tau_0)$,} $\hat{u}(\tau_0,\tau_0)=1$, then we can write a recurrent relation
\begin{equation}\label{B1}
    \begin{aligned}
    \hat{\rho}^{(0)}(\tau)&=\hat{u}(\tau,0)\hat{\rho}(0)\hat{u}^\dagger(\tau,0)\,,\\
    \hat{\rho}^{(1)}(\tau)&=\sum_{j_1}\int_0^\tau d\tau_1 \hat{u}(\tau,\tau_1)\,\hat{l}_{j_1}\,\hat{\rho}^{(0)}(\tau_1)\,\hat{l}_{j_1}^\dagger\hat{u}^\dagger(\tau,\tau_1)\,,\\
    \hat{\rho}^{(n)}(\tau)&=\sum_{j_{n}}\int_0^\tau d\tau_{n} \hat{u}(\tau,\tau_{n})\,\hat{l}_{j_{n}}\,\hat{\rho}^{(n-1)}(\tau_{n})\,\hat{l}_{j_{n}}^\dagger\hat{u}^\dagger(\tau,\tau_{n})\,.
    \end{aligned}
\end{equation}
\noindent Individual terms $\hat{\rho}^{(n)}$ represent the non-Hermitian Hamiltonian evolution of $\hat{\rho}(0)=\sum_k p_k|\phi_k\rangle\langle\phi_k|$ in between $n$ quantum jumps realized by the acting of dissipators $\{\hat{l}_{j_i}\}_{i=1}^n$ at all possible ordered times \mbox{$\{\tau_i\}_{i=1}^n\subset(0,\,\tau)$.}

The MCWF method produces $N$ independent trajectories $|\phi^{(i)}(\tau)\rangle$, where the initial states are sampled from $\hat{\rho}(0)$ eigenstates with respect to their probabilities. In accordance with \eqref{B1}, each trajectory samples a uniformly distributed random number $\xi_1\in (0,\,1)$, which determines the time of the first jump through $\xi_1=S(\tau_1,0)$, where \mbox{$S(\tau_{n+1},\tau_n)=\langle\phi^{(i)}(\tau_n)|\hat{u}^\dagger(\tau_{n+1},\tau_n)\hat{u}(\tau_{n+1},\tau_n)|\phi^{(i)}(\tau_n)\rangle$.} If $\xi_1<S(\tau_f,0)$, then no jump occurs, otherwise at obtained $\tau_1$, a specific event $j_1$ is sampled with respect to the instantaneous rates $\gamma^{(i)}_{j_1}(\tau_1)=\langle\phi^{(i)}(\tau_1)|\hat{l}_{j_1}^\dagger\hat{l}_{j_1}|\phi^{(i)}(\tau_1)\rangle$ and quantum jump $|\phi^{(i)}(\tau_1)\rangle\to\hat{l}_{j_1}|\phi^{(i)}(\tau_1)\rangle/\sqrt{\gamma^{(i)}_{j_1}(\tau_1)}$ is performed. The evolution then proceeds in the same way by sampling $\xi_2$ and taking the normalized post-jump state as the state entering $S(\tau_2,\tau_1)$ until we reach $\xi_n<S(\tau_f,\tau_{n-1})$. The solution to \eqref{Lindblad} can then be approximated by averaging \mbox{$\hat{\rho}(\tau)\approx\frac{1}{N}\sum_{i=1}^N\frac{|\phi^{(i)}(\tau)\rangle\langle\phi^{(i)}(\tau)|}{\langle\phi^{(i)}(\tau)|\phi^{(i)}(\tau)\rangle}\,$.}

In principle, arbitrary accuracy in the estimation of $\hat{\rho}(\tau)$ can be reached by increasing $N$. However, we are primarily interested in evaluating time-dependent expectation values \mbox{$\langle\hat{A}\rangle(\tau)=\textrm{Tr}(\hat{\rho}(\tau)\hat{A})\approx\frac{1}{N}\sum_{i=1}^N\frac{\langle\phi^{(i)}(\tau)|\hat{A}|\phi^{(i)}(\tau)\rangle}{\langle\phi^{(i)}(\tau)|\phi^{(i)}(\tau)\rangle}\,$.} The deviation of obtained expectation values from the exact results can be estimated from the central limit theorem (CLT) as
\begin{equation}
    \sigma_N(\tau)\approx\frac{\sigma_A(\tau)}{\sqrt{N}}=\sqrt{\frac{\langle\hat{A}^2\rangle(\tau)-\langle\hat{A}\rangle^2(\tau)}{N}}\:,
\end{equation}
\noindent where $\hat{A}$ and $\hat{A}^2$ are again averaged over all MCWF trajectories sampled. Therefore, convergence of the method depends on the observable $\hat{A}$ of interest and, in general, the method yields better results for non-local operators. For a more detailed discussion, we refer to \cite{Molmer}.

In Tab. 1, we compare selected aspects of the \eqref{Schro} and \eqref{Lindblad} evolutions at the end of linear-interaction ramps corresponding to the parameters in Fig. \ref{fig:8}. We list the post-protocol mean bosonic populations $\langle\hat{n}\rangle$, mean qubit orientations $\langle\hat{J}_z\rangle$ and down-projected vacuum populations $\tilde{P}_0$. For each of these observables, we evaluated the 'Schrödinger' and 'Lindblad'-acquired mean values, as well as the mean relative errors (MRE) obtained by averaging $N$ independent MCWF trajectories
\begin{equation}\label{B3}
    \sqrt{N}\:\eta_A^{(L)}(N)\approx \frac{\sqrt{\langle\hat{A}^2\rangle(\tau_f)-\langle\hat{A}\rangle^2(\tau_f)}}{\langle\hat{A}\rangle(\tau_f)}\,.
\end{equation}
%\noindent The number of trajectories sampled in each of the protocols is sufficient to estimate the influence of assumed environmental effects on a given observable and the rate of convergence with increasing $N$.
\begin{table}[b!]
\label{Table1}
\caption{Selected observables and MRE \eqref{B3} corresponding to final states obtained by \eqref{Schro} evolutions (indexed by S) and $N$ MCWF trajectories (indexed by L) through different-duration $\tau_f$ interaction ramps matching \mbox{Fig. \ref{fig:8}} model and environmental parameters.}
\begin{tabular}{l|ccc}
\hline
\textbf{Obs./$\bm{(\tau_f , N)}$} & $(10\pi,\, 50\cdot 10^3)$ & $(20\pi,\, 25\cdot 10^3)$ & $(50\pi,\, 2.5\cdot 10^3)$ \\ \hline
$\tilde{P}_0^{(S)}$                     & 0.704                     & 0.485                     & 0.207                      \\
$\tilde{P}_0^{(L)}$                     & $0.703(1)$                & $0.478(2)$                & $0.211(8)$                 \\
$\eta_{P_0}^{(L)}(N)$                   & $0.312/\sqrt{N}$          & $0.803/\sqrt{N}$          & $1.84/\sqrt{N}$            \\ \hline
$\langle\hat{n}\rangle^{(S)}$                         & 2.54                      & 16.7                      & 62.5                       \\
$\langle\hat{n}\rangle^{(L)}$                         & $2.47(1)$                 & $15.5(1)$                 & $60.1(9)$                  \\
$\eta_{n}^{(L)}(N)$                     & $1.30/\sqrt{N}$           & $1.36/\sqrt{N}$           & $0.753/\sqrt{N}$           \\ \hline
$\langle\hat{J}_z\rangle^{(S)}$                       & $-0.4095$                 & $-0.3079$                 & $-0.1302$                  \\
$\langle\hat{J}_z\rangle^{(L)}$                       & $-0.4057(5)$              & $-0.301(1)$               & $-0.122(4)$                \\
$\eta_{J_z}^{(L)}(N)$                   & $0.254/\sqrt{N}$          & $0.684/\sqrt{N}$          & $1.70/\sqrt{N}$           
\end{tabular}
\end{table}

\section{Discussion of suitable experimental parameters}\label{App:C}
In the following, we examine in greater detail the relationship between the model~\eqref{6} and the experimental parameters \eqref{eq:Lambdadelta}-\eqref{eq:lam}. We will justify our choice of values for the relevant quantities by referring to two highly significant works in the field of QPT studies on trapped ions \cite{Pedernales16,Duan21}.

The theoretical approach, introduced in \cite{Pedernales16}, in which the driving fields are detuned from the first sidebands by $\delta_{r,b}$—an approach that allows critical phenomena to be achieved experimentally, as demonstrated in \cite{Duan21} for $^{171}$Yb$^+$—is valid only for $|\delta_{r,b}|\ll\nu$. The values for which \cite{Duan21} report their results correspond to \mbox{$0.022\gtrsim|\delta_{r,b}|/\nu\gtrsim 0.008\,$.} In our case, we assume an axial mode $\nu\approx2\pi\cdot1.0\,\rm MHz$ for the $^{40}$Ca$^+$ ion (based on the setup in ISI Brno \cite{Obsil}), and thus, in the simulations preceding Rabi oscillations in Fig. \ref{fig:8}, we use the values of $|\delta_{r,b}|/\nu\approx0.004\,$. In the expressions for the model parameters, detunings enter into $\lambda$, $\Delta$, and $\varepsilon$. Dynamical manipulation of $\delta_{r,b}$ to adjust $\lambda$ in time is therefore not suitable, as it would result in a change in the system size and the characteristic energy (time) scale. Therefore, the highest possible stability of $\delta_{r,b}$ is desired in order to obtain a fixed value of $\Delta$. The sources and impact of $\delta_{r,b}$ fluctuations, including the relative error $\eta_\Delta\lesssim0.07$, are discussed in \cite{Duan21}. However, the advantage of measuring the projected vacuum population to study the ESQPT structure, as opposed to the phonon population to study ground state QPT, is the significantly lower noise sensitivity in the value of $\Delta$.

The maximum values of the sideband Rabi frequencies within a given protocol are then the experimental parameter of choice for determining the maximum value of the interaction strength $\lambda_f$. The stable ratio of the components in bichromatic driving then ensures a stable interaction regime quantified by $\delta$, which was experimentally set to $|\delta|\lesssim0.02$ in \cite{Duan21}, and it is natural to expect similar fluctuations for asymmetric sideband drive ($\delta\neq0$). In \cite{Pedernales16}, numerical simulations were also performed for the symmetric case $\Omega_1(t)=\Omega_2(t)$, but for the resonant drive of the red sideband, $\delta_r=0$, and thus $\Delta=1$ through \eqref{eq:Del}. The resulting system was then fundamentally limited by finite-size scaling, and therefore it was necessary to simulate even the ground state QPT for the high value of \mbox{$\lambda_f\approx4.53$} achieved for $\eta\Omega_{\rm 1,2}^{\rm (max)}\approx2\pi\cdot4.08\,\rm kHz$ and $\delta_b=-2\pi\cdot1.8\,\rm kHz$. These values are considered for $^{40}$Ca$^+$ and served as our inspiration; however, adjustments for a higher value of $\Delta$ and non-zero $\delta$ were necessary to address the criticality of interest, as we see from Figs. \ref{fig:9} and \ref{fig:10}. In \cite{Duan21} for $^{171}$Yb$^+$, the utility of $\Delta$ scaling was experimentally verified, specifically for the values $5.0\lesssim\Delta\lesssim25.0$ and $\eta\Omega_{\rm 1,2}^{\rm (max)}\approx2\pi\cdot14.2\,\rm kHz$, which correspond to $\lambda_f\approx1.42$, which proved sufficient for ground-state QPT.

Finally, we turn to the expressions \eqref{eq:eps}, \eqref{eq:lam} for $\varepsilon$ and $\lambda$, respectively, and note that scaling the absolute value of detunings affects not only the validity of the RWA and the Rabi frequencies required to maintain a constant $\lambda$, but also the characteristic energy (time) scale $\varepsilon$. This effect is most pronounced in the conversion between laboratory time $t$ and the characteristic time $\tau$ \eqref{TimeConversion}. While the actual duration of the protocol $t_f$ determines the consequences of environmental effects—of which \cite{Duan21} considers only motional decoherence (cf. the overview below \eqref{dissip} and the caption of Fig. \ref{fig:8}) with $0\leq\Gamma_m^{-1}\lesssim 5.5\,\rm ms$—it is $\tau_f$ that determines the unitary evolution itself. In \cite{Pedernales16}, neither dissipation nor decoherence are considered and the hereby hinted interplay is therefore lost; however, in \cite{Duan21}, the relatively large value of $\varepsilon/\hbar\approx2\pi\cdot10.0\,\rm kHz$ is reached, directly impacting the $\tau(t)$ conversion \eqref{TimeConversion}. Thus, the system adapts to QPT on the timescale of $t_f\approx5.0\,\rm ms$, since it converts to $\tau_f\approx 200\pi$. This effect is in part due to the significant system size $\Delta=25$, the scaling impact of which provides the optimal combination for ground state QPT observation (requiring smaller $\lambda_f$) without exposing the system to noticable environmental effects. In our case, we retain a lower value of $\varepsilon/\hbar\approx2\pi\cdot0.98\,\rm kHz$ similar to \cite{Pedernales16}, since the environmental effects are even then negligible for the desired relatively low $\tau_f$ and projected vacuum population measurements. Further scaling of the detunings to achieve larger $\varepsilon$ through \eqref{eq:eps} would reduce the $\lambda_f$ achievable with the same Rabi frequencies through \eqref{eq:lam}.

\end{document}